**From Ultrafast Demagnetization to Ultrafast Spintronics : a 30 years story**

Quentin Remy[1,2] and Stephane Mangin[1,3*]

1)Université de Lorraine, CNRS, Institut Jean Lamour, 54500 Nancy, France
2) Department of Physics, Freie Universität Berlin, 14195 Berlin, Germany
3)Center for Science and Innovation in Spintronics, Tohoku University, Sendai, Japan.

*Stephane.mangin@univ-lorraine.fr

# Abstract

The discovery of femtosecond laser-induced ultrafast demagnetization in 1996 opened a new field—femtomagnetism—in which magnetic order can be quenched on timescales shorter than a picosecond. This seminal observation revealed that angular momentum can be transferred out of the spin system with unprecedented speed, launching intense efforts to disentangle the roles of electrons, phonons, and spins in the non-equilibrium regime. Soon it became evident that ultrafast demagnetization generates spin-flips, spin polarization, magnons and spin currents, providing new channels for angular-momentum flow. These insights laid the foundation for linking femtomagnetism with spintronics. An emblematic breakthrough in this evolution is the *helicity-independent single-pulse all-optical switching* (AOS) observed in rare-earth-transition-metal (RE-TM) ferrimagnets such as GdFeCo. In these alloys, a single femtosecond pulse triggers unequal demagnetization of the antiferromagnetically coupled 3d and 4f sublattices, producing a transient ferromagnetic-like state that enables deterministic reversal through ultrafast angular-momentum transfer. This mechanism, operating at femtojoule-scale energies and without external magnetic fields, establishes RE-TM alloys as benchmark systems for understanding and exploiting angular-momentum flow at the femtosecond timescale. Building on these concepts, the combination of ultrafast optical excitation with spintronic devices has demonstrated deterministic magnetization reversal driven by femtosecond pulses in spin valves and tunnel junctions, including rare-earth-free systems. In these structures, optical demagnetization of one magnetic layer launches a femtosecond spin-polarized current that crosses a non-magnetic spacer and acts as an ultrafast spin-transfer torque, enabling sub-picosecond switching of the adjacent layer. Ultrafast spin injection, acting analogously to spin-transfer torque but operating three orders of magnitude faster, allows reversal of both ferromagnetic and ferrimagnetic layers. Furthermore, optical control of exchange bias, phase-transition-assisted spin-current gating, and hot-electron-driven switching highlight the versatility of ultrafast angular-momentum redistribution in engineered heterostructures. These developments establish *ultrafast spintronics* as a new field at the intersection of femtomagnetism and spin-transport physics, offering pathways toward hybrid photonic-spintronic devices that combine optical writing with magnetic storage. By enabling ultrafast and energy-efficient switching, this emerging paradigm promises scalable technologies for high-speed information processing while raising fundamental questions about angular-momentum transfer in strongly out-of-equilibrium quantum materials.

# 1. Introduction

The nature and relevant timescales of angular momentum have been a central topic of condensed matter physics since the beginnings of quantum mechanics. In 1915, the famous Einstein-de Haas experiment showed that angular momentum could be transferred between the electronic and the lattice degrees of freedom, at least in less than a second[1]. In the 1930's, soon after the concept of spin was introduced, it was clear that one can transfer energy and angular momentum between spins and the lattice, with a dynamic in the second timescale or even longer in paramagnetic salts[2] down to below 100 picoseconds in ferromagnets[3]. This was however just the beginning of the study of paramagnetic[4–7], nuclear magnetic[4,8] and ferromagnetic[9–11] resonance where spin-lattice transfer times cover most of the range from several hours down to around 100 picoseconds. Even with the development of picosecond laser pulses, no new timescales were discovered at first, with first estimations of the spin relaxation times of one nanosecond minimum in Ni[12] and 30 ps minimum in Fe[13], consistent with the known spin-lattice relaxation times. Such pulses had essentially no observable effect on angular momentum. In parallel, angular momentum transport had been studied since 1956 when Torrey extended the Bloch equations to include diffusion terms[14], followed by some of the first works on interfacial spin transport in metals[15–17].

The past fifty years of magnetism research have witnessed repeated redefinitions of how magnetic order can be manipulated, transported, and understood. The discovery of giant magnetoresistance established the foundations of spintronics[18,19], while the birth of femtosecond (fs) laser pulses led to the landmark observation of laser-induced ultrafast demagnetization (UDM) in nickel by Beaurepaire and co-workers in 1996[20]. This work, along with subsequent studies published shortly afterwards[21–25], revealed that magnetic order can collapse within a few hundred femtoseconds. These seminal results marked the emergence of femtomagnetism and demonstrated that spin angular momentum can be redistributed at speeds far exceeding those of conventional spin-lattice relaxation.

In the wake of this discovery, a central challenge was to identify the microscopic pathways responsible for ultrafast angular-momentum transfer. Early phenomenological approaches such as the three-temperature model[20] captured the coupled evolution of electrons, spins, and lattice but could not describe the microscopic origin of the spin loss. Progressively refined frameworks, including notably Elliott-Yafet spin-flip scattering[26–46], non-equilibrium exchange and electron-magnon scattering[33,34,38,47–71], and the spin-transport mechanisms[31,35,36,54,57,59,72–84], revealed that femtosecond demagnetization involves both *local* (spin-orbit mediated) and *non-local* (transport-mediated) redistribution of angular momentum. These ideas established that optical excitation does not simply heat the electrons but also launches ultrafast spin-polarized currents that propagate across tens of nanometers, thereby forming a natural bridge between femtomagnetism and spintronics.

In parallel, major experimental advances deepened our understanding of these processes. Time-resolved magneto-optical measurements and element-resolved probes such as X-ray Magnetic Circular Dichroism (XMCD) and high-harmonic spectroscopy[85–87] showed that different magnetic sublattices demagnetize at different rates[88], revealing transient ferromagnetic-like states in ferrimagnets[89]. Time resolved X-ray diffraction showed evidence that angular momentum ends up in the lattice[90] via the generation of polarized phonons[91]. Terahertz-emission spectroscopy further demonstrated that femtosecond demagnetization emits ultrashort spin currents detectable as THz radiation[92–102]. Together, these techniques established a microscopic picture in which angular momentum flows through intertwined electronic, spin, and lattice degrees of freedom on ultrafast timescales[103–107].

A major conceptual milestone was the discovery of all-optical switching[108], which demonstrated that laser excitation can not only quench magnetization but also deterministically reverse it without any magnetic field. Four distinct types of all-optical switching (AOS), governed by different underlying mechanisms, can be distinguished. Helicity-dependent switching[109,110] (AO-HDS) requires multiple pulses[111], and the light helicity influences both domain nucleation and domain-wall motion[112]. A second class is a precessional switching (AO-PS), recently observed in specific rare-earth-based heterostructures[113–116], where a single pulse induces in-plane magnetization reorientation through a precessional pathway rather than the conventional toggle mechanism. A third class deals with inherently non-thermal processes, where magnetization is reversed via the coherent excitation of a phonon mode[117,118] or d-d transitions[119]. A fourth class is helicity-independent single-pulse switching[89,108,120–124], (AO-HIS) which provides ultrafast, deterministic reversal with a single femtosecond pulse. In rare-earth-transition-metal (RE-TM) ferrimagnets such as GdFeCo, this process arises from ultrafast and unequal demagnetization of the antiferromagnetically coupled 3d and 4f sublattices, producing a rapid angular-momentum exchange[89]. Operating at femtojoule energy scales, this mechanism remains one of the most efficient magnetic-writing processes known and will be detailed in this review.

The last decade has seen the emergence of a unified field for *ultrafast spintronics*, in which femtosecond optical excitation is integrated with spin-transport architectures. In spin-valve heterostructures, demagnetization-driven spin currents act as ultrafast spin-transfer torques[79,125,126] capable of reversing an adjacent magnetic layer on sub-picosecond timescales[127–132]. Similar effects have been achieved in magnetic tunnel junctions, where fs-spin-polarized carriers tunnel through MgO barriers and induce deterministic reversal within a few picoseconds[133]. These spin currents can also be used to promote or hinder helicity-independent single-pulse switching of ferrimagnets, thereby eliminating the toggle-switching property of this single-pulse switching in favor of a fully deterministic reversal[134]. Hot-electron-driven switching[135] further demonstrated that femtosecond spin-current pulses can be generated without direct optical absorption in the magnetic layer, opening pathways compatible with device integration[136,137]. Alternatively, when the spacer in such spin-valve heterostructures is thin enough, the Ruderman-Kittel-Kasuya-Yosida coupling between the ferrimagnetic and ferromagnetic layers can be strong enough such that the helicity-independent single-pulse switching of the ferrimagnet drives the reversal of the ferromagnet[138], although with a slower dynamics compared to spin current-induced switching[139,140]. Similar ideas allow a fully optical and ultrafast control of exchange bias in IrMn/GdCo bilayers[141].

These developments illustrate a coherent evolution: from femtosecond quenching of magnetization[20–25], to the identification of microscopic channels of angular-momentum flow[40,63,85,89,91,106,142], to ultrafast switching in complex heterostructures[127,130,131], and ultimately to the emerging vision of photonic-spintronic hybrid devices operating at THz rates[143].

In this article, we trace this thirty-year development from ultrafast demagnetization to ultrafast spintronics. Section 2 revisits the microscopic mechanisms of ultrafast demagnetization, ultrafast angular-momentum flow through spin currents, magnons, and THz emission as well as their conceptual impact. Section 3 discusses single-pulse all-optical switching in RE-TM heterostructures and identifies unifying physical principles. Section 4 reviews ultrafast spintronic devices, including hot-electron switching, ultrafast spin-transfer torque, and tunnel-junction reversal. Section 5 offers an outlook on hybrid photonic-spintronic architectures and emerging opportunities for ultrafast magnetic technologies.

## 2. Ultrafast Demagnetization (Beaurepaire *et al.* and beyond)

### 2.1 Discovery and phenomenology

The first direct evidence that magnetic order could collapse on sub-picosecond timescales came from the pioneering experiment of Beaurepaire et al.[20], who observed that the remanent magnetization of nickel is quenched within a few hundred femtoseconds following excitation by a femtosecond laser pulse. As shown in Fig. 1, the transient longitudinal Magneto-Optic Kerr Effect (MOKE) signal of a Ni/MgF$_2$ thin film exhibits an abrupt drop immediately after excitation, with the magnetization reaching a minimum within ≈300 fs before recovering toward equilibrium on a several-picosecond timescale. This behavior was radically different from the spin-lattice relaxation processes previously associated with hundred-picosecond demagnetization and spin relaxation in general, and demonstrated that magnetic order can be altered far more rapidly than any known thermal pathway.

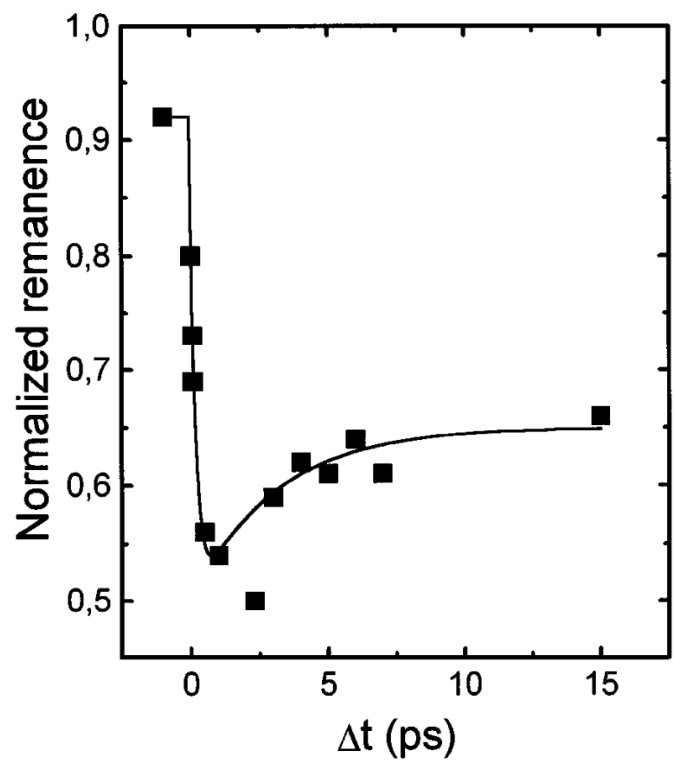

*Figure 1.* Longitudinal MOKE signal of a Ni(22 nm)/MgF$_2$(100 nm) film as a function of pump-probe delay[20]. The pump excitation is a 7 mJ/cm$^2$ pulse with a duration of 60 fs duration. The signal is normalized to the static signal without pump. The sharp drop within a few hundred femtoseconds provides the first direct evidence of ultrafast demagnetization, while the picosecond recovery reflects electron-spin-lattice equilibration. The solid line is a guide to the eye.

This observation marked the birth of *femtomagnetism* by demonstrating that ultrashort optical excitation drives magnetic materials into a strongly non-equilibrium regime in which angular momentum can be redistributed on femtosecond timescales. In metallic ferromagnets, a femtosecond laser pulse initially creates a highly non-thermal electronic population, which relaxes toward a quasi-thermal distribution through electron-electron scattering within approximately 10–50 fs[144–146]. During this early non-equilibrium window, the electronic

structure—including spin-dependent occupations and potentially exchange splitting[34,44], see also[48,50,64]—is transiently modified, enabling exceptionally efficient angular-momentum transfer between spin, orbital, and transport degrees of freedom. Spin-orbit coupling, ultrafast hot-electron transport, and collective excitations such as magnons provide parallel channels for this redistribution, which are either ineffective or entirely absent under equilibrium conditions.

In the decades following this seminal work, ultrafast demagnetization was observed in a wide variety of materials—including elemental 3d ferromagnets[21–25], rare-earth metals[147–152], rare-earth-transition-metal ferrimagnets with coupled 3d-4f sublattices[153], semiconductors[38,154], Heusler alloys and half-metals[155–159], insulators (although through slower spin-lattice coupling, which we do not consider further in this review) [160,161], and multilayer heterostructures[28,73,83,162–165]. These observations established sub-picosecond quenching as a universal response of magnetically ordered solids driven far from equilibrium[103,106].

Today, ultrafast demagnetization is recognized as a cornerstone phenomenon for understanding ultrafast angular-momentum flow in magnetic materials. It provides the physical foundation for several key developments discussed in this review, including single-pulse all-optical switching and the generation of femtosecond spin currents used in terahertz spintronics[143].

**2.2. Electron-spin-lattice non-equilibrium: beyond the three-temperature picture**

The first phenomenological framework proposed to describe ultrafast demagnetization was the three-temperature model (3TM)[20], in which electrons, spins, and phonons are treated as coupled reservoirs characterized by distinct temperatures and exchanging energy through effective coupling constants. A schematic representation of this approach is shown in Fig. 2. The 3TM successfully captures the hierarchy of characteristic timescales observed experimentally, namely the rapid heating of the electronic system, the subsequent evolution of the spin system, and the slower equilibration with the lattice.

However, by construction, the model assumes that each subsystem remains internally thermalized at all times and focuses exclusively on energy exchange. As a consequence, it cannot describe the strongly non-equilibrium electronic distributions and angular-momentum-selective processes that dominate during the first few hundred femtoseconds after excitation. In particular, the 3TM does not explicitly account for microscopic mechanisms such as electron-magnon scattering, spin-orbit-mediated angular-momentum transfer, or non-local spin transport, and therefore underestimates the role of direct electron-spin coupling and the pathways through which angular momentum is redistributed to the lattice[63].

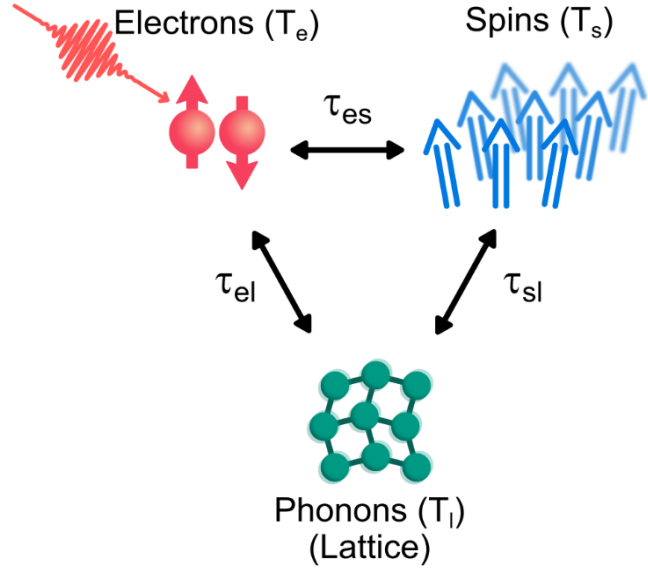

*Figure 2.* Schematic representation of the three-temperature model[20]. Electrons, spins, and phonons are described as coupled subsystems, each characterized by an effective temperature, $T_e$, $T_s$, and $T_l$, respectively. Energy is exchanged between these reservoirs through phenomenological coupling/time constants, while each subsystem is assumed to remain internally thermalized at all times. A femtosecond laser pulse only acts on the electronic reservoir.

Subsequent experimental and theoretical studies have shown that the electronic system must instead be viewed as a set of interacting subsystems with distinct roles in both energy and angular-momentum redistribution[31–33,40,54,55,61,63,70,85,91,146,166–170]. Their interplay defines the strongly non-equilibrium regime of femtomagnetism.

One of the first microscopic mechanisms considered is spin-orbit-mediated spin-flip scattering, often referred to as Elliott-Yafet scattering[5,7], which converts spin angular momentum into orbital motion that is subsequently transferred to the lattice[26,29,40,91,171–174]. This class of processes encompasses several microscopic mechanisms since spin-orbit coupling implies that many scattering events can contribute to spin dissipation[71,172]. More recently, additional spin-dissipation channels requiring spin-orbit coupling have been proposed as dominant contributors to demagnetization[63,175]. These processes differ from conventional Elliott-Yafet scattering in that they do not necessarily involve single-particle spin flips.

In parallel, the different mean free paths and velocities of majority and minority carriers generate superdiffusive hot-electron currents that transport angular momentum away from the excited region[35,81–83,176–178]. This mechanism, illustrated in Fig. 3, involves majority and minority carriers propagating across nanometric distances immediately after excitation. These ultrafast spin currents can cross interfaces into adjacent layers and constitute the electronic basis of terahertz spintronics. They are directly detected in THz-emission experiments such as those shown in Fig. 4.

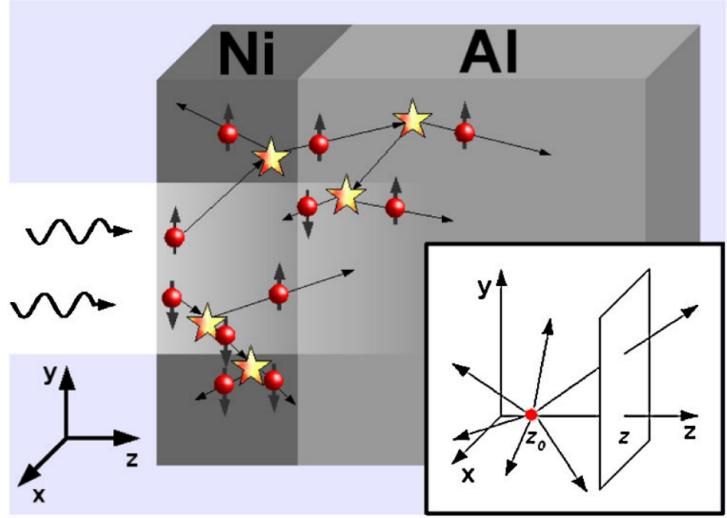

*Figure 3.* *Superdiffusive spin transport following femtosecond laser excitation[81]. Majority and minority hot electrons propagate with different velocities and inelastic mean free paths, producing a transient spin-polarized current that removes angular momentum from the excited region.*

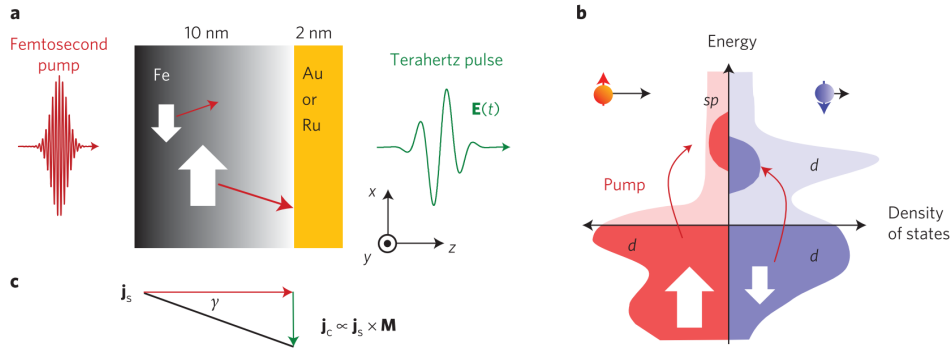

*Figure 4.* *Terahertz emission from a ferromagnet/normal-metal bilayer[92]. (a) Femtosecond demagnetization drives an ultrafast spin current into the adjacent heavy metal, where spin-orbit coupling converts it into an in-plane charge pulse that radiates a broadband THz transient. (b) The excitation transforms slow majority-spin d electrons (red) into fast sp electrons, thereby launching a spin current towards the gold or ruthenium cap layer. (c) Illustration of the relationship between the spin current, charge current and the spin Hall angle $\gamma$.*

Another important, though still debated, contribution arises from the collapse of exchange splitting in the non-thermal electronic distribution[34,44,146]. A reduction of exchange splitting has been observed in photoemission measurements of some materials. However, the interpretation remains debated: in some cases the observations have been explained in terms of band mirroring[48,50], while in others such a collapse was not observed, for instance in iron[64]. Further work is therefore required to determine whether this mechanism plays a non-negligible role in ultrafast demagnetization.

More recent studies have uncovered additional ultrafast channels that further enrich this picture. Attosecond photoemission experiments have demonstrated optical intersite spin transfer (OISTR)[73], a process in which optical excitation directly shifts spin-polarized charge between neighboring atoms on attosecond to few-femtosecond timescales. Such ultrafast interatomic spin displacement represents one of the fastest angular-momentum transfer pathways known[78,179]. While initially observed in iron-nickel alloys[180], the associated dynamics have also been interpreted in terms of inhomogeneous magnon generation[77,181–183].

Complementing these microscopic perspectives, the concept of a transient spin voltage[31,143,184] provides a unified thermodynamic description of ultrafast angular-momentum redistribution. In this framework, optical excitation combined with electron-magnon scattering[33,63] generates a difference in electrochemical potential between majority and minority carriers, producing a transient spin accumulation that drives both local demagnetization and the emission of spin currents. This description naturally connects femtosecond magnetization quenching with THz-emission experiments.

The relative importance of these mechanisms depends strongly on material composition[135,185], interface transparency[163,186], and the nature of the magnetic order[104]. Ferrimagnets exhibit particularly rich dynamics due to the coexistence of 3d and 4f sublattices with distinct exchange interactions[187,188]. Element-resolved XMCD measurements (Fig. 5) reveal that these sublattices demagnetize on different timescales, leading to ultrafast inter-sublattice angular-momentum transfer and the formation of transient ferromagnetic-like states[89,120]. These phenomena constitute the microscopic basis of helicity-independent all-optical switching discussed later in this review.

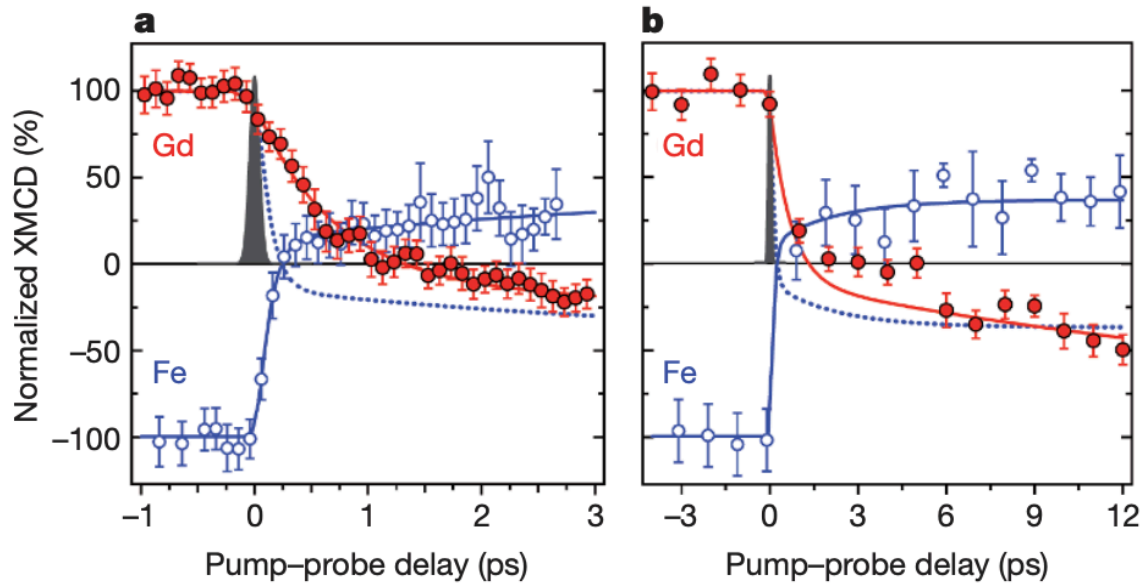

*Figure 5. Element-specific XMCD traces in GdFeCo[89]. The 3d and 4f sublattices demagnetize on distinct timescales, revealing ultrafast inter-sublattice angular-momentum transfer and the formation of a transient ferromagnetic-like state.(a) and (b) correspond to the same dynamics on different timescales.*

Taken together, these results demonstrate that ultrafast demagnetization does not originate from a single microscopic process but rather from the combined action of several strongly non-equilibrium mechanisms. Spin-orbit-mediated spin-flip scattering[26,29,40,171–174], magnon generation[33,34,48,50,61–64], superdiffusive hot-electron transport[35,81–83,176–178] and the transient collapse of exchange splitting[34,44,146] all contribute to the redistribution and dissipation of angular momentum. In addition, collective excitations such as short-wavelength magnons mediate angular-momentum transfer within and between magnetic sublattices[89,120,180–183].

Recent theoretical work suggests that electron-magnon scattering may provide a unifying microscopic framework for both ultrafast demagnetization and spin-current generation[31,33,47,57,189]. In this picture, transient spin and magnon chemical potentials drive angular-momentum exchange between itinerant electrons and the magnon reservoir. When an adjacent layer acts as an efficient spin sink, the interfacial spin current becomes directly proportional to the temporal derivative of the magnetization[57,63]. The universal sub-picosecond quenching observed in a wide variety of magnetic materials[20,103,104] therefore reflects a complex interplay of local and non-local angular-momentum transfer processes, whose relative importance depends on material composition, interface transparency, spin-mixing conductance, and the efficiency of spin dissipation in the surrounding layers.

## 2.3. Where does the angular momentum go? Conservation, flow, interfaces, and ultrafast spin pumping

The sub-picosecond collapse of magnetization raised the long-standing question of how angular momentum is conserved during such a rapid process. Modern theory and experiment now agree that ultrafast demagnetization does not destroy angular momentum but redistributes it among several strongly coupled reservoirs. Spin-orbit coupling (SOC) enables the transfer of angular momentum from the spin to the orbital sector on femtosecond timescales[40,171]. Electron-phonon and electron-electron scattering as well as spin-lattice interactions provide other channels that transfers angular momentum to the lattice[103], in the presence of SOC. Transport mechanisms also contribute: superdiffusive spin-polarized hot electrons can leave the excited region and carry angular momentum across interfaces within a few tens of femtoseconds[81,92]. It must be noted that while the consequences of spin transport on magnetization dynamics in general is of the greatest importance[125–127,130,131], it is not the main ingredient of UDM and only partially contribute to it[31,35,190], mainly since UDM is also observed in single thin ferromagnetic layer where transport cannot exist. Collective excitations redistribute angular momentum internally within the magnetic system[61,151] which is essential to understand the dynamics of magnetic alloys as noted above.

Although many theoretical attempts have been made to date to theoretically simulate ultrafast demagnetization (notably but not limited to references[26,32,33,38,40,44,46,47,61,62,65,148,171,173,191–194]), there is still no theoretical framework able to reproduce all the characteristic features of UDM. Likewise, the general physical understanding of UDM is still limited and seems to strongly depend on the details and nature of the samples such as the magnetic material of choice, the various layer thicknesses and the material choice for capping layers and substrates. A significant step in this direction was recently made thanks to ultrabroadband THz emission spectroscopy[31,63,195] which studies the low fluence excitation regime, also known as the linear regime, for single ferromagnetic layers without optical absorption gradients and sandwiched between insulating materials. In this regime, it was realized that magnetization dynamics at ultrashort timescale does not merely consists of UDM and (fast and slow) magnetization recovery, as always believed so far and as predicted by most models, especially temperature models[104]. Indeed, before UDM starts, magnetization dynamics is already happening via the concomitant generation of magnons (decreasing the electronic angular momentum) and Stoner excitations (those increasing the electronic angular momentum) via electron-magnon scattering, while conserving both the total electron+magnon energy and angular momentum i.e. without demagnetization. This is illustrated in Figure 6. Figure 6b illustrates the essence of the phenomenological model behind the experimental observations of Rouzegar *et al.*[63]. Mathematically, the linear regime is described by a set of coupled linear ordinary differential equations, similar to the (linearized) three-temperature model. The unknowns of this system of equations are typically thermodynamic variables[46] (or generalized thermodynamic variables[31]), based on the fact that at low fluences, the system is driven only slightly away from equilibrium. From an experimental point of view, the general good practice is to find the minimum set of such thermodynamic variables that can describe this non-equilibrium situation. We note that more strongly non-thermal situations can in principle be described by increasing the number of thermodynamic variables[46] such as what is done for non-thermal electronic distributions[196,197], phonons[198,199] and magnons[61]. In the case of electrons, several studies tend to indicate that non-thermal distributions do not play a role[162,197,200–204], even in the non-linear regime (for the studied fluences)[205]. In the case of the work of Rouzegar *et al.*[63], a key ingredient was to consider that not only energy but also angular momentum must be exchanged between the individual baths of the system. Even though this makes sense from a theoretical point of view[206], experiments focusing on local dynamics was only focused on energy transfers[104] while angular momentum transfer was mostly considered in spin transport. In the case where angular momentum transfer needs to be explicitly accounted for, one also needs to include chemical potentials as unknowns of the system of equations[46]. Then the smallest set of thermodynamic variables contains the electron, magnon and phonon temperatures $T_e$, $T_m$ and $T_p$ respectively, as well as the spin independent and spin dependent (also known as spin voltage or spin accumulation) electronic chemical potentials $\mu_e$ and $\mu_{e,s}$, and the magnon chemical potential $\mu_m$. This set can be written as a vector $\xi = (\delta\mu_e, \mu_{e,s}, \delta T_e, \mu_m, \delta T_m, \delta T_p)$ where $\delta$ indicates the variation with respect to the initial equilibrium state which has an electronic chemical potential $\mu_e^0$, a common temperature $T_0$ and the magnon chemical potential and spin accumulation are zero. Note that contrary to the three-temperature model, both electrons and magnons carry energy and angular momentum. In fact, one should see electrons and magnons as two different types of electronic excitations[33,63,207,208]. Accordingly, the set of thermodynamically conjugate densities is $\chi = (N_e, M_e, E_e, M_m, E_m, E_p)$ where $N$, $M$ and $E$ stand for particle density, angular momentum density and energy density and the subscripts $e$, $m$ and $p$ stand for electrons, magnons and phonons respectively. From the dispersion relation of each reservoir, obtained from *ab initio* calculations or experiments, one can obtain a linear relationship between $\xi$ and $\chi$[63]

$$\chi = \chi^0 + \Xi\xi \qquad (1)$$

Where $\chi^0$ is the equilibrium value of $\chi$. The dynamics of the system is then phenomenologically described in the relaxation time approximation by

$$\dot{\xi} = \sum_\alpha \Gamma_\alpha \xi \qquad (2)$$

$$\Gamma_\alpha = \frac{1}{\tau_\alpha}(A_\alpha \Xi - B_\alpha)^{-1} B_\alpha \qquad (3)$$

With $\Gamma_\alpha$, $A_\alpha$ and $B_\alpha$ some 6 by 6 matrices and $\tau_\alpha$ the scattering time associated with the microscopic interactions considered between each bath. These interactions between any two baths are fully defined, in this framework, by whether the interaction allows a transfer of energy or angular momentum or both. By constraining energy, number of electrons and angular momentum to be conserved together with the appropriate equilibrium conditions (equalities shown in Figure 6b), the matrices $A_\alpha$ and $B_\alpha$ are fully defined[63]. The solutions of equation (2), with appropriate initial conditions (typically an elevated electron temperature), are linear combinations of decaying exponentials whose characteristic times are given by the eigenvalues of $\sum_\alpha \Gamma_\alpha$. Two such characteristic times are always infinity based on the conservation of energy and electronic number (there is no eigenvalue associated with the conservation of angular momentum since the phonon angular momentum density is not included in $\chi$). As noted above, this model relies on the assumption of small fluences and each reservoir is assumed in equilibrium at all times. If the latter assumption is not verified, it is in principle possible to increase the dimension of the system of equations (2) as previously mentioned. In the case of the work of Rouzegar et al.[63], this was found to not be necessary as the solution of equation (2) gives the minimum number of decaying exponentials needed to fit the THz emission data (Figure 6a). However, the scattering times $\tau_\alpha$ should always be understood as effective time since studies with a shorter time resolution may reveal additional hidden dynamics, just like the demagnetization time from the Stoner or temperature models[31,32] is an effective time which can be derived from equation (2).

The conclusions of this analysis are illustrated in Figure 6b. In a first step, after the laser pulse deposits energy in the electrons, energy and angular momentum is exchanged between electrons and magnons at the 10 fs timescale via electron-magnon scattering. In this step, both magnons and Stoner excitations are generated while the total angular momentum of the system is still (approximately) constant. The characteristic time of this dynamics spans from around 4 fs in Co to 17 fs in $Ni_{50}Fe_{50}$ and scales with the inverse of the sample Curie temperature. During this step, not only the magnon temperature increases, but the magnon chemical potential decreases. At the same time, the spin accumulation increases. The latter two facts are crucial, since they allow the total angular momentum to be conserved. This is absent from the three-temperature model which is then unable to reproduce the fact that magnons can be excited without the system demagnetizing. UDM itself happens only at the 100 fs timescale, via exchange of angular momentum between magnons and phonons. This happens because, in the presence of SOC, the magnon chemical potential must vanish in equilibrium. Microscopically, this corresponds to something like three-magnon scattering[175,209]. This can also be modeled by considering the spin-orbit coupling of d electronic states[171]. While the latter effect is already considered in ab initio simulations like time dependent density functional theory (TDDFT) simulations[173,210,211], it is not clear whether it is the same mechanism as three-magnon scattering or in general an angular momentum transfer between magnon excitations and phonons, since TDDFT typically includes a small subset of magnons if any[173], does not include phonons beyond the Ehrenfest dynamics[212] and does not conserve angular momentum[213]. Surprisingly, it is found that contrary to the usual assumption[38,47,54,132,167], the spin dissipation from the electrons (single particle states) towards the lattice is negligible. The reason is that when a realistic magnon dispersion relation is taken into account, the magnon spin susceptibility is much larger than the electron spin susceptibility, because it typically requires much less energy to excite a magnon compared to a Stoner excitation. Then, the electronic bath prefers to equilibrate with the magnon bath instead of the lattice at ultrashort timescales[63]. The latter fact is typically missed by the standard implementation of electron-magnon scattering in the mean-field approximation (see reference [63] for a detailed discussion). Finally, on longer timescales, magnetization recovers due to electron-phonon coupling. It must be noted however that this remagnetization is the combined effect of three interactions. Other effects, such as the role of spin heat accumulation[84], energy exchange between magnons and phonons[214,215], dynamic exchange splitting[44] and a direct laser induced spin accumulation[31] are found to have a negligible effect on the dynamics itself while it may affect its amplitude. We note however that this may no longer hold in the non-linear regime or at ps timescales.

These ideas should extend to the non-linear regime, although there is no existing complete theory yet. However, in our opinion, the framework of Dusabirane et al.[33] is the closest to such a theory and the small clusters simulations of Töws and Pastor[171] seem to include the same physics with a reduction in the spin-spin correlation function (magnon generation) happening after a few tens of femtoseconds while demagnetization itself happens at the few hundreds of femtoseconds timescale. The spin dynamics at these short timescales in the non-linear regime is also still experimentally unexplored. In that regard, we also note that the calibration procedure of ultrabroadband THz emission spectroscopy relies on the linearity of the dynamics. Thus, from an experimental perspective, studying the non-linear regime at ultrashort timescales will either require a significantly more complex analysis of THz emission data, or the use of different experimental methods compatible with short pulse durations and high fluences such as attosecond MCD[73,78].

The generation of magnons at ultrashort timescales was observed before[67,69], it is known that UDM must happen via the generation of magnons[34,48,50,64] (the extent of the contributions of Stoner excitations at longer time delays is still debated, mostly because of the results of reference[34]) and there has been a few experimental evidences that there must be an energy transfer between electrons and magnons[55,56,146] although in some cases this was attributed to either spin-flip scattering or spin transport[146]. However, it was only later found that this energy transfer must be accompanied by an angular momentum transfer, and that both happen via electron-magnon scattering at a timescale one order of magnitude shorter than previously anticipated[63]. It is worth noting that the Stoner excitations generated during electron-magnon scattering are fundamentally different from those of conventionally discussed spin-flip scattering mechanisms (with SOC), since in the former case the electronic spin increases while in the later it decreases[33].

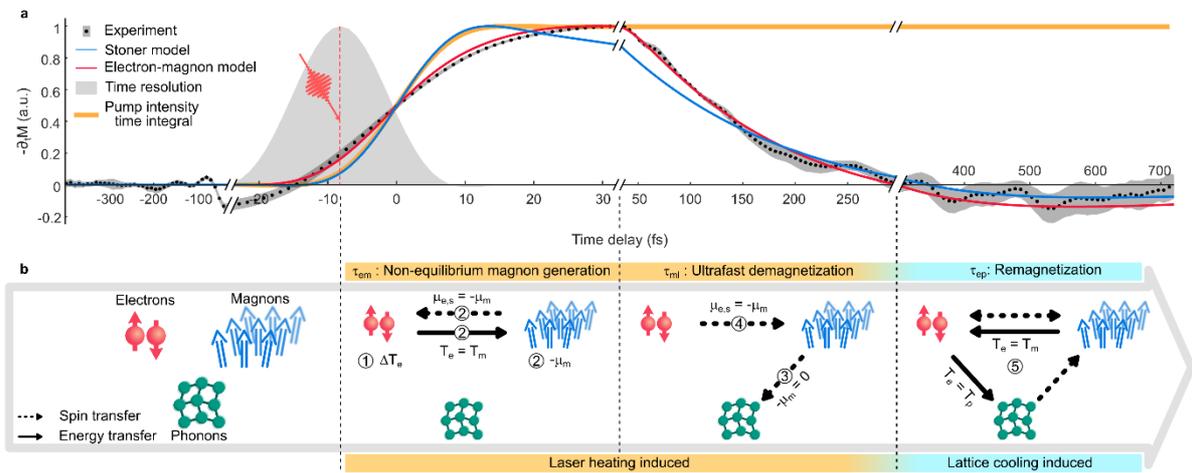

*Figure 6. (a) Experimentally determined demagnetization rate $-\partial_t M$ for permalloy (dotted line) using THz emission spectroscopy[63]. The yellow curve is the time integral of the effective pump-pulse intensity envelope (light grey shaded area). The blue curve is the best fit from the Stoner model (or equivalently a temperature model) while the red curve shows the result of the phenomenological electron-magnon model. The dark grey area around the data points represents the uncertainty. (b) Scenario of magnetization dynamics according to the phenomenological electron-magnon model. The numbers indicate the various steps of the spin dynamics while the equalities next to each arrow indicates the equilibrium conditions.*

The previous considerations can be used to directly connect ultrafast spin dynamics and spintronics. Indeed, in spintronics, currents (and mostly spin currents) are driven by various gradients. In the general non-equilibrium setting, currents are driven by gradients of the particle (electron, magnon, phonons, etc.) distribution functions of the system, which are different from the equilibrium distribution functions in general[50,196,216–218]. Sufficiently close enough to equilibrium, or in the linear regime, one can characterize these distribution functions in terms of their statistical moments only which are temperatures and chemical potentials for Fermi-Dirac and Bose-Einstein distribution functions. If we consider a typical ferro(ferri)magnet/non-magnet bilayer (where the ferro(ferri)magnet may or may not be metallic), where the thickness of each film is sufficiently thin (typically much less than the optical absorption length) and the non-magnet is a good spin sink, the spin current at the bilayer interface reads[31,84,219,220]

$$j_s = j_s^e + j_s^m,$$
$$j_s^e = \frac{1}{2\pi}\frac{g_{\uparrow\uparrow}g_{\downarrow\downarrow}}{g_{\uparrow\uparrow}+g_{\downarrow\downarrow}}\mu_{e,s} + \gamma_\uparrow(T_{e,\uparrow} - T_{e,N}) - \gamma_\downarrow(T_{e,\downarrow} - T_{e,N}),$$
$$j_s^m = \frac{\mathrm{Re}\,g_{\uparrow\downarrow}}{\pi s}\left[D_m^1 \mu_m + D_m^2 \frac{T_m - T_{e,N}}{T_0}\right]. \qquad (4)$$

The (interfacial) spin current contains an electronic $j_s^e$ as well as a magnonic $j_s^m$ contribution. Possible contributions of phonons[221] or electronic orbital angular momentum[222–227] are not considered in this formula and in this review. The spin conductance matrix $g$ over the two-dimensional spin space $(\uparrow, \downarrow)$ is given by the interface area $A$ and reflection matrices $r_\sigma$ ($\sigma = \uparrow, \downarrow$) via $g_{\sigma\sigma'} = \mathrm{tr}(1 - r_\sigma r_{\sigma'}^\dagger)/A$[228]. The first term of the electronic contribution, proportional to the spin accumulation is the main effect usually invoked to explain ultrafast spintronics[31], especially THz emission in spintronic THz emitters[31] and ultrafast magnetization reversal[128,130]. From a theoretical perspective, beyond the restricted domain of validity of equation (4), it has been described using different formalisms[57,59,76,81,177,193,217,229]. Intuitively, it corresponds to a spin current originating from an excess of spin in one region of space compared to another (and comes here from a "gradient" of spin accumulation at the bilayer interface). The remaining two terms correspond to the spin dependent Seebeck effect (SDSE)[84,230] where we used a slightly more generalized expression to consider possible different temperatures ($T_{e,\uparrow}$ and $T_{e,\downarrow}$) between both electronic spin species[31,84]. $T_{e,N}$ is the electron temperature of the non-magnetic layer and $\gamma_\uparrow$ and $\gamma_\downarrow$ are coefficients which can be related to phenomenological quantities (spin dependent conductivities and Seebeck coefficients together with relevant boundary conditions[84]) as well as microscopic ones including, but not exclusively, the diagonal element of the spin conductance matrix[31]. The electronic contribution as a whole does not depend on the off-diagonal elements of the spin conductance matrix. Experimentally, the SDSE was found to have only an impact at long timescales i.e. above a picosecond[31,230]. $j_s^e$ is obviously non-zero only in fully metallic bilayers. The magnon contribution on the other hand does not depend on the diagonal elements of the spin conductance matrix and can be non-zero even when one layer is insulating[231,232]. It depends only on the real part of the spin mixing conductance since there is here no coherent spin precession (all the magnetization dynamics is longitudinal). The common pre-factor also depends on the spin density $s$. Inside the brackets, the coefficients $D_m^n$ depend only on the magnon density of states and the equilibrium temperature $T_0$ via[63] $D_m^n = \int d\varepsilon \left(-\frac{df^0}{d\varepsilon}\right)\varepsilon^n \nu_m(\varepsilon)$ where $f^0$ is the equilibrium Planck distribution at temperature $T_0$ and $\nu_m$ is the magnon DOS. Both terms come from interfacial electron-magnon scattering[47,231]. The second term is the spin Seebeck effect (SSE)[231,233]. Equation (4) does not include the effect of carrier multiplication in the non-magnetic layer, which is important for the SSE[231,233]. Carrier multiplication tends to decrease the amplitude of this contribution of the spin current as well as slow down its dynamics overall[195,231]. The effect of carrier multiplication on the spin dynamics of (metallic) ferromagnets has not been observed[63] and its role for the SSE most likely comes from the interfacial nature of the effect. The first term of the magnon contribution has been previously suggested as an important contribution for ultrafast interfacial spin currents[57,189]. Since this term does not depend on the non-magnetic layer (its "magnon chemical potential" is always zero), it is not impacted by carrier multiplication. Also, because the magnon chemical potential and the spin accumulation have an almost identical (normalized) dynamics[63], it is difficult to separate this contribution of the spin current from the electronic spin current driven by the spin accumulation. Estimates based on realistic magnon DOS[63] indicate that the chemical potential driven magnonic spin current should be around an order of magnitude smaller than the spin accumulation driven electronic spin current. Nevertheless, it is still a challenge to confirm these predictions experimentally. The observation of ballistic magnon transport in $Fe_{20}Ni_{80}$/NiO/Pt trilayers[232] supports but does not confirm the possibility of a interfacial spin current driven by the magnon chemical potential. Note again that these expressions are only valid for thin bilayers in the linear regime (or sufficiently close to equilibrium). They convey, however, all the standard physics of ultrafast spintronics.

It may be convenient to choose an appropriate name for the magnonic spin current term driven by the magnon chemical potential. We think it makes sense to call it (ultrafast) spin pumping, since it scales with the spin mixing conductance, as in standard spin pumping[234]. In previous works which did not consider the magnon chemical potential[59], spin pumping was used to refer to the combined effect of electron-magnon scattering and spin transport due to a gradient of spin accumulation. In general, electron-magnon scattering gives rise to both a spin accumulation and a magnon chemical potential because of the simultaneous creation of two different types of spin excitations, Stoner excitations and magnons, respectively. The spin transport resulting from each type of spin excitations is not necessarily the same in amplitude, as shown by equation (4), even though they both scale with the demagnetization rate $-\partial_t M$[63]. The first term of the electronic spin current may be referred to as (ultrafast) superdiffusive spin transport (even though magnons may also experience superdiffusive transport) since historically superdiffusive transport was associated with conduction electrons (single electron states) only. Note that superdiffusive should be here understood as including ballistic and diffusive transport as special cases. Then, one can say that ultrafast spin injection is driven by electron-magnon scattering (giving the common $-\partial_t M$ behavior[57,72,235]) and happens via both superdiffusive spin transport and spin pumping.

It has been a key conceptual advance to realize that ultrafast demagnetization can drive an ultrafast spin-current emission process. Unlike conventional spin pumping at GHz frequencies, which originates from coherent magnetization precession[236], and standard spin diffusion[17,237], this ultrafast regime is governed by the rapid, non-adiabatic temporal variation of the magnetization during demagnetization. The proportionality between the spin current amplitude and the time derivative of magnetization has been directly evidenced by terahertz-emission spectroscopy experiments, where the emitted THz field from FM|NM bilayers was shown to follow the ultrafast magnetization dynamics via spin-to-charge conversion mechanisms such as the inverse spin Hall effect, thereby providing a direct experimental link between femtosecond demagnetization and ultrafast spin transport[87,61]. The principle of a spintronic terahertz emitter is shown in figure 7.

Terahertz-emission spectroscopy provided clear and direct evidence for this mechanism[31,63]. In a FM|NM bilayer, femtosecond laser excitation triggers demagnetization in the ferromagnet, as described above, generating a spin current that flows into the adjacent normal metal following equation (4). Spin-orbit coupling in the normal metal converts this spin current into a transverse

charge current through the inverse spin Hall effect, producing a sub-picosecond THz electromagnetic transient[92,93,96–100]. In references[31,63] (see also[101]) the temporal shape of the emitted THz field was shown to follow the time derivative of the magnetization, demonstrating that ultrafast demagnetization produces a spin-current pulse on the same timescale.

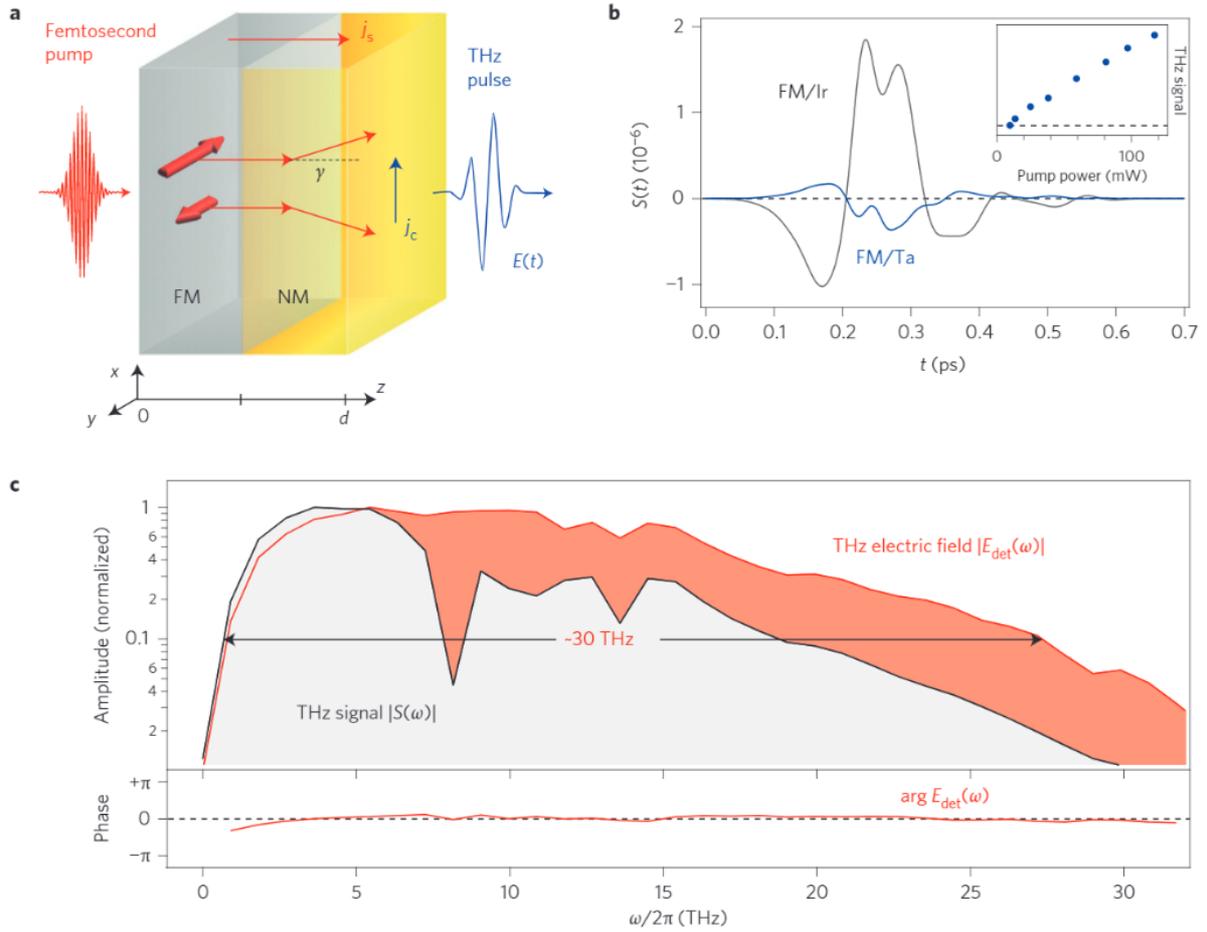

*Figure 7. Principle of a spintronic terahertz emitter and typical emission from[93]. (a) A femtosecond laser pulse generates a spin-polarized current in a FM|NM bilayer; spin-orbit coupling in the NM layer converts it into an ultrafast transverse charge current, producing a THz electromagnetic transient. (b) Example of electro-optic THz signal S(t) from capped CoFeB films and its scaling with pump power (inset). (c) Corresponding Fourier spectra showing broadband emission extending to tens of terahertz and an approximately flat spectral phase characteristic of a near-Fourier-limited THz pulse.*

This result was further consolidated where systematic studies of metallic trilayers such as W/CoFeB/Pt established that the THz signal provides a quantitative measure of the ultrafast spin current flowing across the interface[93,238]. These studies firmly established ultrafast spin transport as a fundamental consequence of femtosecond demagnetization.

The non-local nature of angular-momentum flow in magnetic multilayers was already demonstrated by Malinowski et al.[163], who showed that ultrafast demagnetization in one Co/Pt multilayer accelerates or slows the demagnetization of another, depending on their relative magnetization directions, when the two are separated by a metallic spacer that transmits hot-electron spin currents. When the spacer was changed to an insulating material, the effect disappeared entirely as sketch in figure 8. These experiments provided the first direct demonstration that angular momentum can be transferred from one magnetic layer to another on femtosecond timescales via ultrafast spin currents.

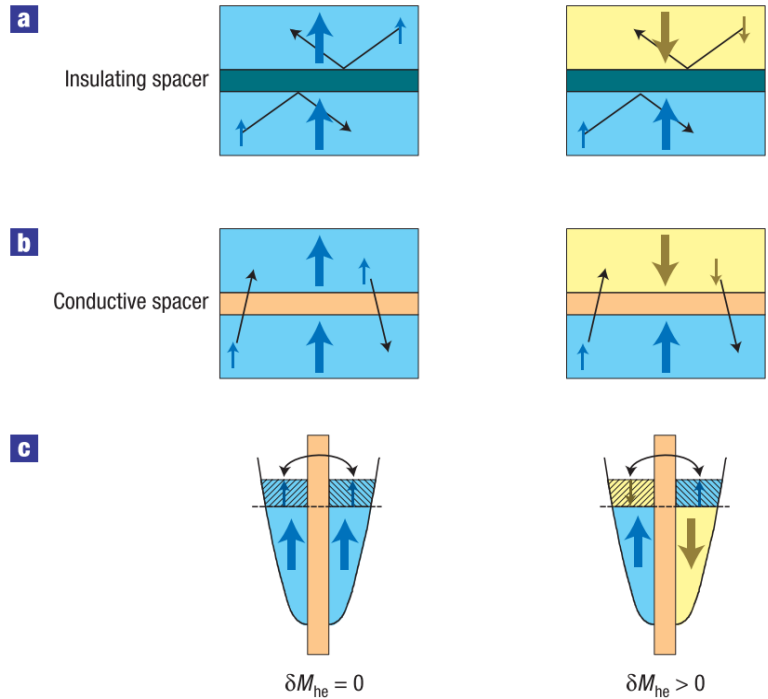

*Figure 8.* Hot-electron transport and interlayer spin-angular-momentum transfer in Co/Pt multilayers from ref[163] (a) When the two magnetic stacks are separated by an insulating spacer, hot electrons excited by the femtosecond pulse are localized, preventing interlayer transport of spin angular momentum. (b) With a metallic spacer, hot electrons propagate across the spacer and transfer spin angular momentum between the two Co/Pt multilayers. (c) Schematic majority-spin parabolic bands of the two multilayers. The quantity $\delta M_{he}$ denotes the demagnetization component arising from hot-electron-mediated spin transfer.

A clear demonstration of interlayer angular-momentum transfer controlled by interface transparency is provided by the time-resolved MOKE measurements shown in Fig. 9, adapted from Malinowski *et al*[163]. In these experiments, the ultrafast demagnetization dynamics of antiferromagnetically coupled Co/Pt multilayers was investigated for different spacer materials and magnetic configurations. When the two magnetic stacks are separated by an insulating NiO spacer, the normalized demagnetization traces measured in the parallel (P) and antiparallel (AP) configurations are nearly identical, indicating that spin-polarized hot electrons remain localized within each magnetic layer. In contrast, for a thin metallic Ru spacer, a pronounced difference between the P and AP configurations is observed: the AP state exhibits both a faster demagnetization and a larger demagnetization amplitude. This behavior directly reflects the presence of an additional angular-momentum dissipation channel enabled by spin-conserving hot-electron transport across the Ru spacer. The comparison between NiO and Ru spacers thus demonstrates that interlayer spin-current transmission can substantially modify both the rate and efficiency of ultrafast demagnetization, highlighting the key role of interface transparency in governing femtosecond angular-momentum flow in magnetic multilayers.

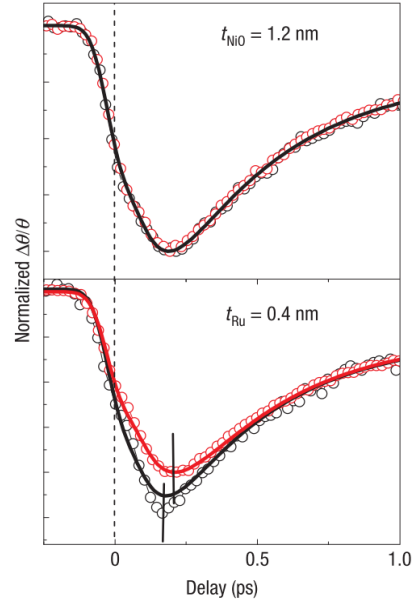

*Figure 9. Normalized TR-MOKE measurements. measured in the parallel (red) and antiparallel states in [Co/ Pt]$_4$/ Co/ S/Co/[Pt / Co]$_4$ heterostructure with S= 1.2 nm of NiO (top figure) and 0.4 nm of Ru (bottom figure) . Curves are normalized at 3 ps, after electron thermalization. The solid lines are fits to the data*

More recent THz-emission studies[239–249] underline that the magnitude, polarity, and bandwidth of these ultrafast spin currents are highly sensitive to interfacial properties such as spin-mixing conductance, layer thickness, chemical intermixing, and the choice of capping layer. Interface engineering has therefore become central to the design of materials for ultrafast spin transport.

In ferrimagnetic systems, the unequal demagnetization of the 3d and 4f sublattices enables angular momentum to be exchanged internally before being dissipated to the lattice or transported away[89,127,128,130,235]. These exchange-mediated flows generate transient ferromagnetic-like states and play a key role in helicity-independent all-optical switching, as discussed in Section 3.

Overall, ultrafast demagnetization does not violate angular-momentum conservation. Instead, it reflects a rapid (local and non-local) redistribution among electronic, spin, lattice and sublattice reservoirs. The emergence of ultrafast spin pumping and superdiffusive spin transport, supported by terahertz-emission experiments, provides a direct and quantitative link between femtosecond magnetization dynamics and ultrafast spintronic functionality.

## 2.4. Critical slowing down and ultrafast spin cooling

A fundamental limitation of thermally driven magnetization reversal is the phenomenon of critical slowing down (CSD), which emerges when the magnetization amplitude approaches zero in the vicinity of the Curie temperature[40,46,65,68,130,250–256]. The simplest approaches to understand this regime are mean-field descriptions which predict a collapse of the exchange-induced energy splitting between spin states, leading to identical transition rates and a vanishing longitudinal magnetization dynamics, irrespective of the electronic temperature[40,54,130]. As a result, the magnetization remains transiently trapped near zero, even though this state is thermodynamically unstable. This bottleneck is intrinsic to models in which the spin system evolves exclusively through thermal coupling to an external bath[257].

Ultrafast optical excitation can circumvent this limitation by coupling the spin subsystem to non-thermal angular-momentum reservoirs[130]. A direct experimental demonstration is provided by the injection of femtosecond spin-polarized hot electrons generated in an adjacent GdFeCo layer and transferred into a Co/Pt multilayer, enabling magnetization reversal within approximately 400 fs—well beyond the speed limit imposed by purely thermal dynamics[130]. In this scenario, the injected spin current acts as an external source of angular momentum that lifts the degeneracy of the spin levels near $m \approx 0$, thereby breaking the rotational symmetry of the paramagnetic state and restoring finite transition asymmetry[130].

As shown in Figure 10, within the two-level mean-field framework, this effect is captured by an additional angular-momentum term $\Delta J_{ext}$ to $\Delta E$, which induces an effective energy splitting $\Delta E_{ext}$ between spin states. As long as $\Delta E_{ext} \neq 0$, the transition rates remain unequal and critical slowing down is avoided. Depending on the sign of the injected spin polarization, the spin system can undergo ultrafast spin heating or ultrafast spin cooling, corresponding respectively to an increase or decrease of the effective spin temperature on sub-picosecond timescales. The bipolar temporal structure of the injected spin current can even result in successive reversal events separated by only a few hundred femtoseconds[125].

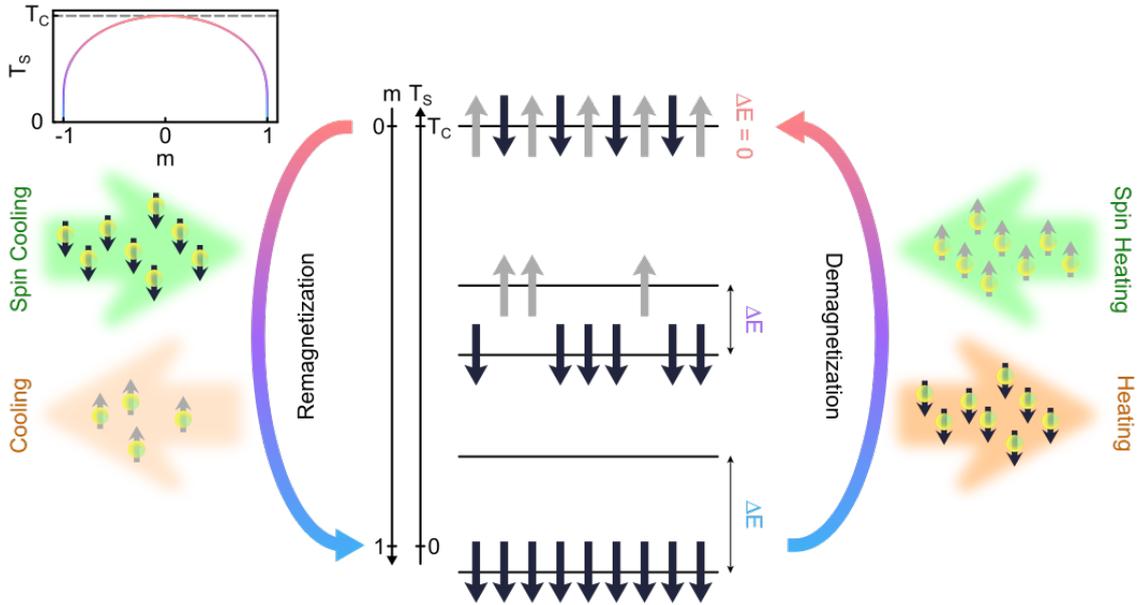

*Figure 10. Two-level mean-field picture of critical slowing down and its suppression by ultrafast spin cooling and heating. The magnetization is represented by the population imbalance between two spin states separated by an energy $\Delta E$ proportional to the normalized magnetization $m$. Near $m = 0$, the collapse of $\Delta E$ leads to identical transition rates and critical slowing down. An external source of angular momentum, such as an injected spin accumulation, introduces an additional splitting $\Delta E_{ext}$ that breaks the degeneracy and restores finite magnetization dynamics. Depending on the sign of the injected angular momentum, the spin system undergoes ultrafast spin heating or spin cooling, enabling sub-picosecond magnetization reversal. Adapted from Ref.[130].*

Beyond enabling ultrafast switching, critical slowing down can also be exploited in a reversed manner to stabilize long-lived magnetization states in systems that are otherwise antiferromagnetic like $FePS_3$[258]. Taken together, these results demonstrate that both critical slowing down and its non-equilibrium lifting through finite spin accumulation hold significant potential for ultrafast spintronics and energy-efficient magnetic switching[130,258].

## 2.5. Unifying perspective and impact of ultrafast demagnetization

Ultrafast demagnetization is now understood as the collective outcome of strongly non-equilibrium interactions among electronic, spin, and lattice degrees of freedom on femtosecond timescales. This framework reveals a hierarchy of dynamical processes: rapid electronic thermalization within tens of femtoseconds, angular-momentum redistribution occurring over approximately 5–200 fs, electron-phonon equilibration on the picosecond timescale, and lattice relaxation extending over several tens of picoseconds. The interplay of these coupled channels governs the response of magnetic materials driven far from equilibrium and sets the fundamental timescales of femtomagnetic dynamics[63,104].

A major conceptual consequence is that ultrafast spin dynamics provides the microscopic foundations for ultrafast spintronics (see the previous section). The transient spin-voltage/accumulation generated during demagnetization and the associated superdiffusive spin transport and spin pumping act as sources of femtosecond spin currents, enabling ultrafast spin-transfer torque[72,125,126,217], hot-electron-driven switching[127–131,135,136,259], and broadband terahertz spin-to-charge conversion[92,93,95–101]. These processes establish a unified link between femtomagnetism and ultrafast spin devices.

As emphasized in the recent Guest Editorial in ref[143] ultrafast demagnetization remains an open problem in several respects, including the quantitative role of interfaces, the interplay between magnon and phonon transport, and the development of predictive multiscale models that treat electronic, orbital, and lattice angular momentum on equal footing. Continued advances in element-specific X-ray and high-harmonic probes, now approaching sub-10-fs resolution, promise to resolve and confirm the earliest microscopic steps of angular-momentum transfer and to push femtomagnetism toward the attosecond regime.

The same ingredients also underpin all-optical switching in rare-earth-transition-metal ferrimagnets[167,260–263]. Unequal demagnetization of 3d and 4f sublattices[89,264], exchange-driven angular-momentum flow[121,124,167,265], and short-lived spin-current bursts[99,235,266–270] together enable helicity-independent single-pulse reversal[89,108,120,121,271–273] as detailed in the next section.

# 3. Single pulse All Optical switching in rare-earth transition-metal (RE-TM) heterostructure and MnRuGa

## 3.1. Gd based alloys

The discovery that a single femtosecond pulse can reversibly switch GdFeCo without any magnetic field marked a defining milestone in femtomagnetism[89,108,120]. In these ferrimagnetic alloys, the antiferromagnetically coupled FeCo (3d) and Gd (4f) sublattices relax on distinct timescales: the transition-metal spins collapse within approximately 100 fs, whereas the Gd moments respond more slowly, in around 400 fs as show in Figure 5[89]. This temporal asymmetry brings the transition-metal sublattice close to complete demagnetization precisely when it is affected by the exchange mediated angular-momentum transfer generated by the ultrafast demagnetization of the Gd sublattice. The resulting angular-momentum flow drives the system through a transient ferromagnetic-like state, enabling deterministic reversal within a few picoseconds[89,120,121,265,274] A systematic mapping of switching phase diagrams in $Gd_x(FeCo)_{100-x}$ alloys established the influence of composition, thickness, and pulse duration. As presented in Figure 11, Wei *et al.*[275] showed that the switching fluence increases approximately linearly with pulse duration, while the multidomain threshold remains nearly constant, producing a triangular operational window whose width is maximized for short pulse duration and near the magnetization compensation composition, $x_{comp} \approx 25\%$[275]. Atomistic spin simulations confirmed that electron overheating (not phonon heating[170]) is the key driver of reversal, in agreement with earlier observations that transient electron temperatures govern the non-equilibrium exchange between sublattices[120,135].

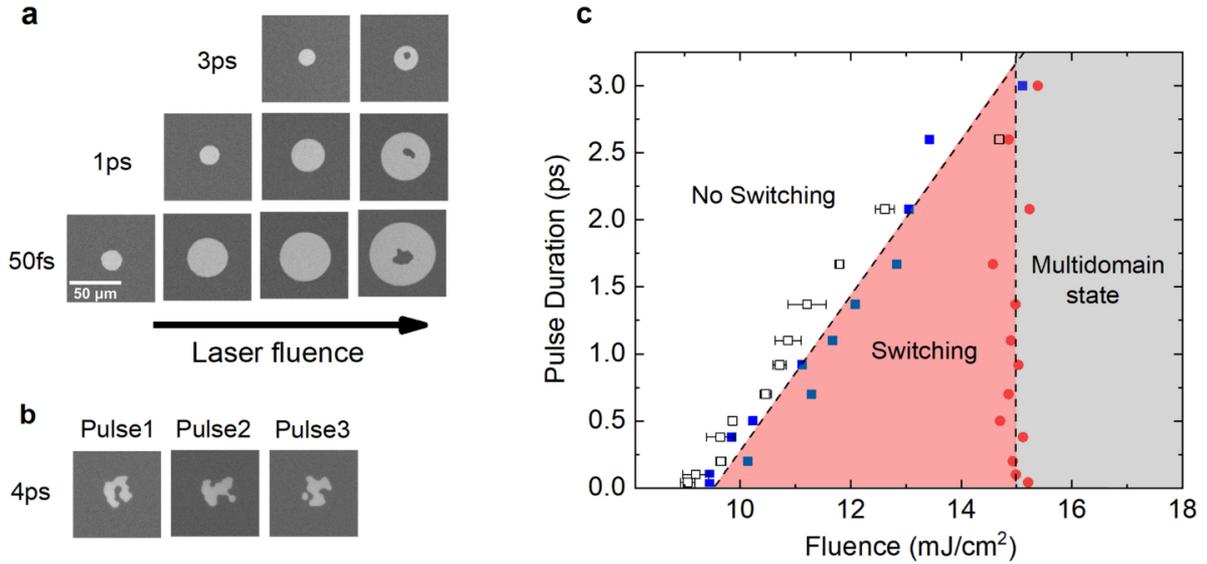

*Figure 11. Magneto-optical images and all-optical helicity-independent switching (AO-HIS) state diagram for a 20 nm Gd24(FeCo)76 film[275]. (a) MOKE images of the magnetic configuration after exposure to a single linearly polarized laser pulse with pulse durations of 50 fs, 1 ps, and 3 ps, for increasing fluence (9.5–15 mJ/cm²). Above the switching threshold fluence $F_{SW}$, deterministic AO-HIS is observed and the switched area expands with fluence (scale bar: 50 μm). (b) MOKE images recorded after a single linearly polarized pulse with a longer duration of 4 ps at 17 mJ/cm² (Pulse1-Pulse3), showing fully demagnetized/multidomain patterns rather than deterministic reversal. (c) AO-HIS state diagram reporting the switching fluence $F_{SW}$ (open black squares and filled blue squares) and the multidomain fluence $F_{MD}$ (filled red circles) as a function of pulse duration. Filled blue squares correspond to $F_{SW}$ defined when the diameter of the switched area reaches ~10 μm, while open squares are extracted using the fitting procedure proposed by Liu et al.[276]. The spatial full-width at half-maximum (FWHM) of the laser beam is ~70 μm.*

A practical limitation of observing AO-HIS in perpendicularly magnetized Gd-TM films is the onset of thermally activated multidomain formation, which collapses the deterministic switching window at high fluence and/or long pulse duration. This limitation is made explicit in Figure 12 (adapted from Fig. 3 of Ref.[170]), where the switching threshold fluence $F_{SW}$ increases monotonically with pulse duration $\tau_{las}$ but remains essentially independent of the Cu heat-sink thickness $t_{Cu}$, whereas the multidomain threshold $F_{MD}$ increases strongly with $t_{Cu}$ and displays a saturating dependence on heat-sink thickness for a given $\tau_{las}$. As a consequence, enlarging $t_{Cu}$ widens the operational window $F_{SW} < F < F_{MD}$ and pushes the maximum pulse duration[277] for which AO-HIS can be observed (defined by $F_{SW} = F_{MD}$) to substantially longer values. Verges *et al.*[170] interpreted $F_{MD}$ as the onset of a thermally driven loss of deterministic reversal, occurring when insufficient cooling at longer delays after electron-phonon equilibration leads either to complete demagnetization or to dipolar-field-driven breakup into domains, rather than to an intrinsic failure of the ultrafast switching mechanism itself. This interpretation is strongly supported by a recent theoretical work[278], who used multiscale micromagnetic simulations including heat transport in the full stack to reproduce the experimentally observed No Switching (NS), Switching (SW), and MultiDomain (MD) regions. Their analysis shows that the NS→SW boundary is governed mainly by the intrinsic ultrafast thermal dynamics within the GdFeCo layer and is therefore only weakly affected by $t_{Cu}$, whereas the SW→MD boundary is controlled by the cooling rate at longer times: when both sublattices remain fully demagnetized for too long above $T_C$, the system relaxes into a random multidomain state, while a thicker Cu underlayer enhances heat extraction and stabilizes recovery into the switched single-domain state. Beyond thermal management using metallic heat sinks, an additional route to suppress multidomain formation is to reduce dipolar fields by moving away from perpendicular anisotropy: in-plane magnetized GdCo alloys largely avoid maze-domain stabilization, enabling single-pulse AO-HIS over a much broader composition/thickness range, as demonstrated in Ref.[185].

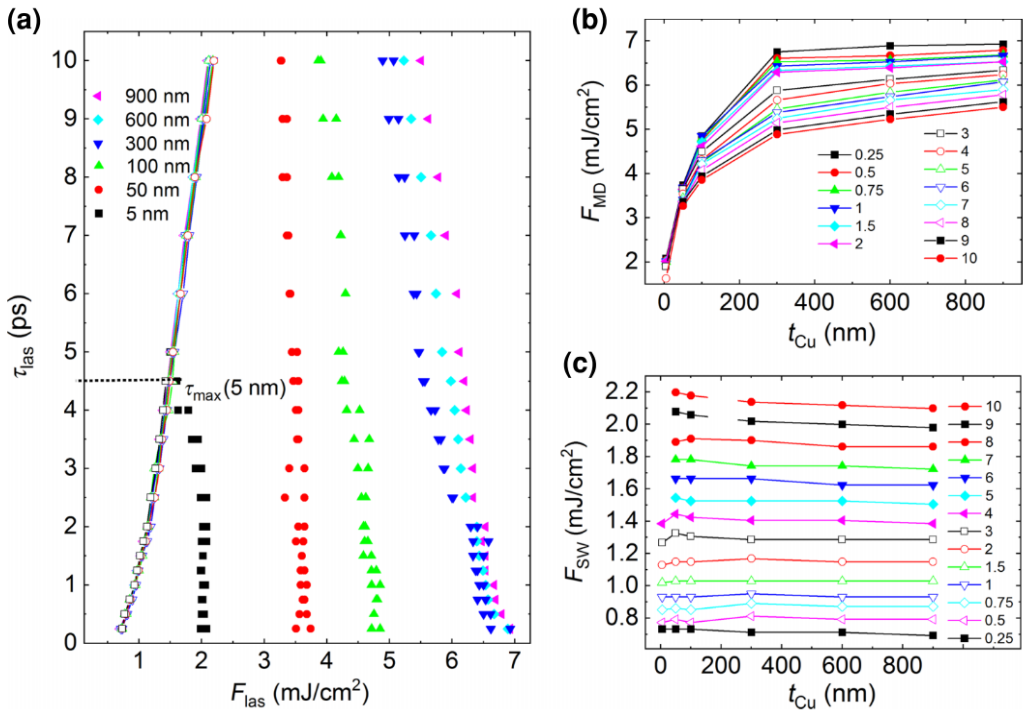

*Figure 12. (adapted from Ref.[170]). (a) Switching fluence $F_{SW}$ (open symbols, solid lines) and multidomain fluence $F_{MD}$ (filled symbols) as a function of pulse duration $\tau_{las}$, for different Cu heat-sink thicknesses $t_{Cu}$ (legend). The dashed line indicates the maximum pulse duration $\tau_{max}$ for which AO-HIS is observed for $t_{Cu} = 5$ nm, defined by $F_{SW}(\tau_{max}) = F_{MD}(\tau_{max})$. (b) $F_{MD}$ versus $t_{Cu}$ for several $\tau_{las}$ values (in ps, legend), showing a strong increase and a saturating dependence with heat-sink thickness. (c) Same as (b) for $F_{SW}$, evidencing that the switching threshold is essentially independent of $t_{Cu}$ over the explored range.*

Early AO-HIS experiments (including the state-diagram approach) were predominantly carried out in Gd-based ferrimagnets with strong perpendicular magnetic anisotropy (PMA) and compositions close to magnetization compensation, which naturally led to the widespread assumption that "near compensation + strong PMA" were intrinsic prerequisites for single-pulse AO-HIS. However, a growing body of results indicates that these conditions are often practical, not fundamental: the dominant limitation in PMA systems is the tendency to break into multidomain states driven by dipolar fields and delayed thermal processes, which collapses the deterministic switching window at high fluence/long pulses. A direct way to suppress this extrinsic limitation is to reduce stray-field-driven domain formation by moving to in-plane magnetization. This is illustrated in Fig. 13 where longitudinal MOKE images demonstrate deterministic single-pulse AO-HIS in in-plane $Gd_XCo_{100-X}$ over a broad concentration range (well beyond a narrow vicinity of compensation) and for multiple thicknesses: panels (a–c) show reproducible toggle-like reversal across Co-dominant and Gd-dominant alloys, while panel (d) summarizes the composition dependence of the switching and demagnetization thresholds, together with a damage threshold, highlighting that AO-HIS can persist far from compensation when dipolar-field-driven maze domains are avoided.

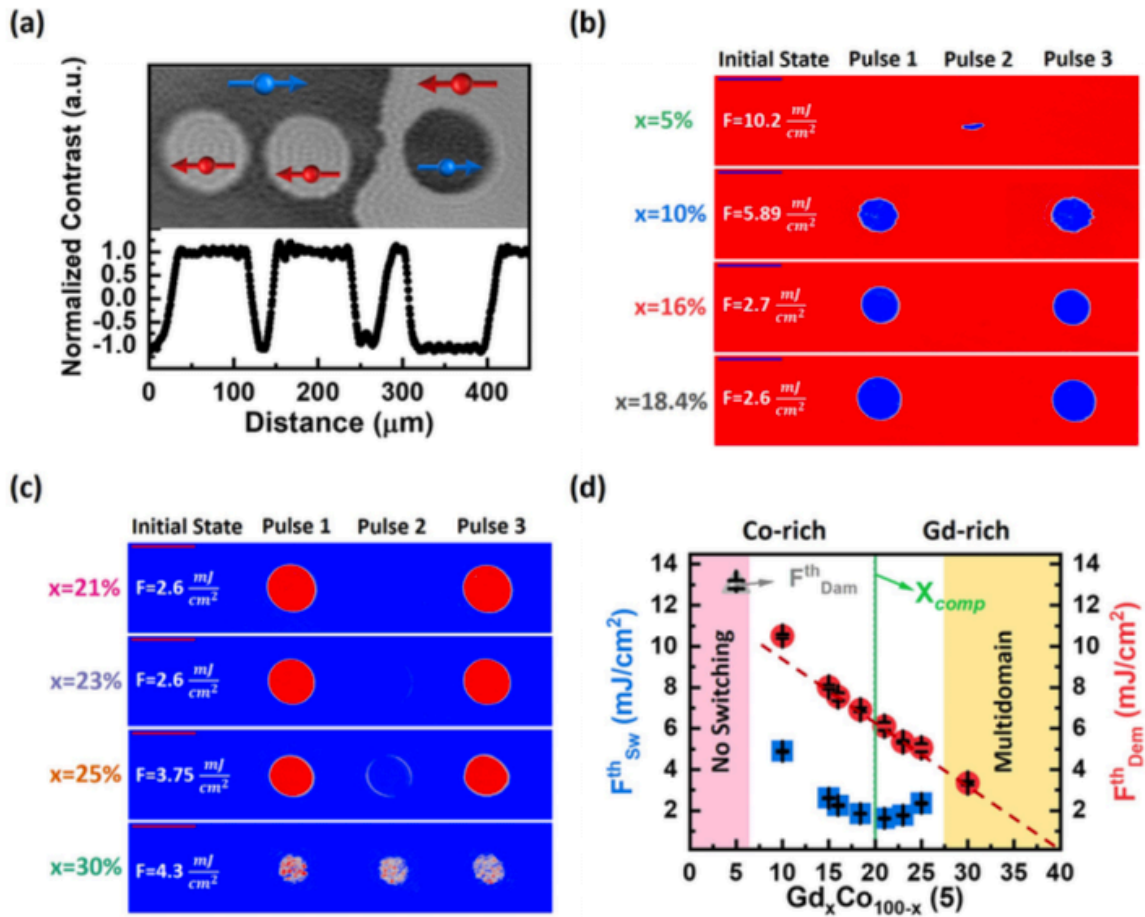

*Figure 13.* (adapted from Ref[185]). *AO-HIS in in-plane magnetized $Gd_XCo_{100-X}$ alloys measured by longitudinal MOKE microscopy with single 150-fs laser pulses. (a) Kerr images and normalized contrast profiles after three single shots on a 5-nm $Gd_{15}Co_{85}$ film starting from a two-domain initial state; arrows indicate the Co-sublattice magnetization direction. (b,c) Magneto-optical contrast after single-pulse excitation for Co-dominant and Gd-dominant*

*compositions (same position), demonstrating deterministic AO-HIS over a wide concentration range. (d) Switching threshold $F_{th}^{Sw}$, demagnetization (multidomain) threshold $F_{th}^{Dem}$, and damage threshold $F_{th}^{Dam}$ as a function of Gd concentration, showing that AO-HIS can be observed far from compensation when dipolar-field-driven multidomain formation is suppressed by in-plane anisotropy.*

Complementary routes to mitigate multidomain formation and relax "compensation constraints" include (i) reducing the effective dipolar cost through thinner or interface-engineered stacks and (ii) tuning magnetic parameters by rare-earth substitution. In particular, Fig. 14[279] emphasizes that even a very small amount of interfacial Gd "dusting" in an otherwise Co/Pt-based PMA system can trigger robust single-pulse AO-HIS—i.e., switching can be realized extremely far from bulk compensation, pointing to an interface-local (or strongly nonuniform) ferrimagnetic functionality and underscoring the role of interface engineering.

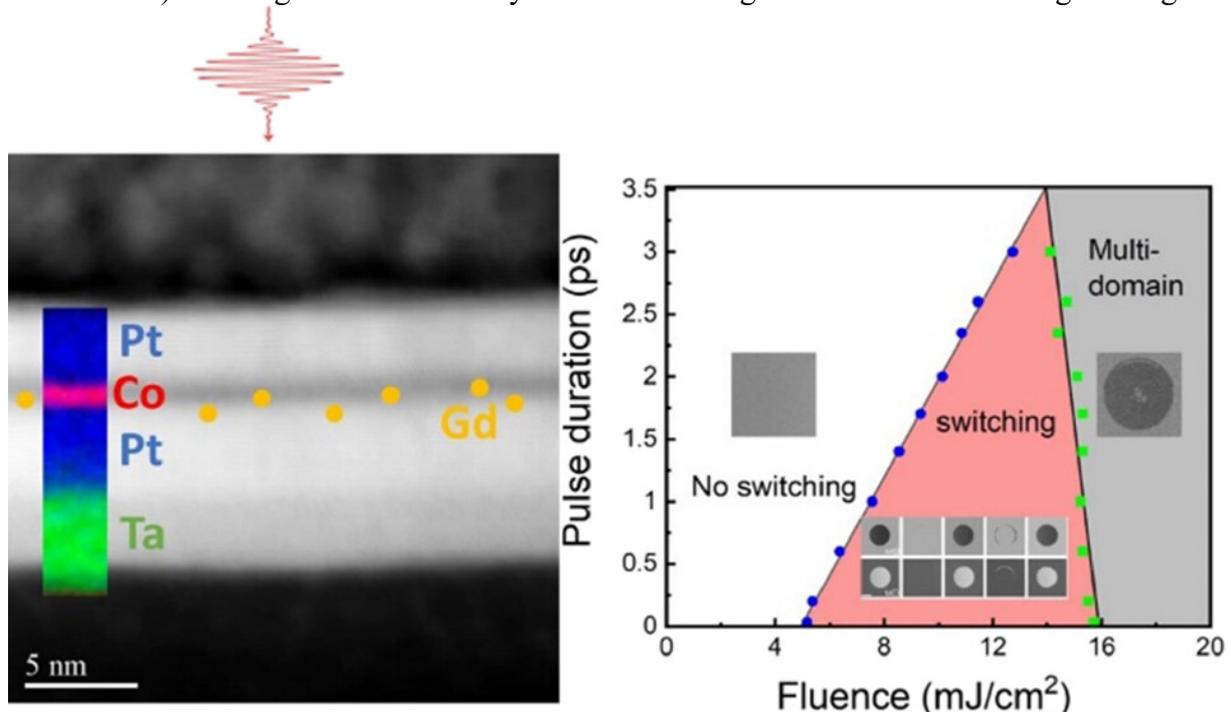

***Figure 14.*** *(adapted from Ref.[279], cover image). Graphical illustration of interface-enabled single-pulse AO-HIS in a Pt/Co/Pt-based heterostructure by introducing an ultrathin "Gd dusting" layer at one Pt/Co interface, highlighting that deterministic ultrafast switching can be achieved with a very small rare-earth content through interface engineering.*

It is important to emphasize that AO-HIS in RE-TM ferrimagnets is often described as a spatially uniform and purely local process, governed exclusively by ultrafast angular-momentum transfer between antiferromagnetically coupled sublattices. However, the results presented in Fig. 15 reveal that this description is incomplete. In a 9.4 nm Gd25Co75 layer, magnetization reversal is strongly inhomogeneous along the film depth due to asymmetric excitation and cooling imposed by the surrounding layers. After laser excitation, two regions with opposite magnetization directions transiently coexist, separated by a highly mobile boundary that propagates across the layer thickness. As a result, AO-HIS is governed not only by local ultrafast demagnetization, but also by the picosecond propagation of this transient boundary. Both the switching speed and the final magnetic state are therefore determined by the interplay between local reversal near the hotter side of the film and non-local front propagation toward the colder side. This demonstrates that nanoscale depth inhomogeneities must be explicitly considered in realistic descriptions of AO-HIS in magnetic heterostructures.

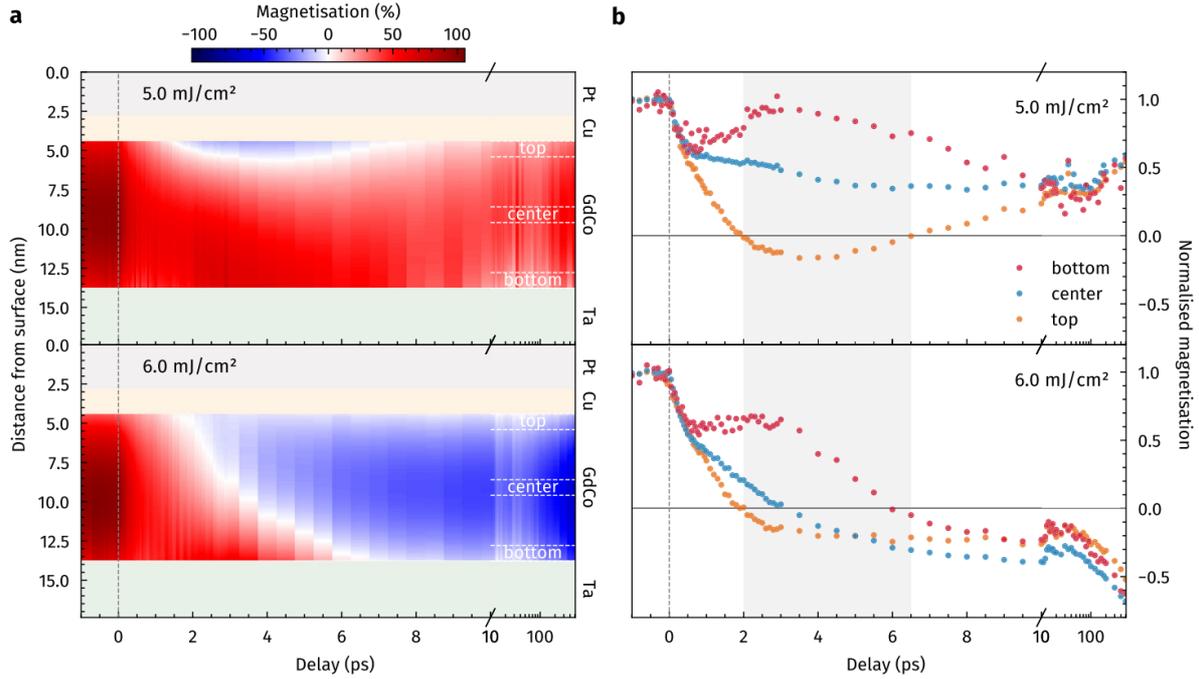

*Figure 15. Depth-resolved AO-HIS dynamics in a 9.4 nm $Gd_{25}Co_{75}$ layer, adapted from Fig. 4 of Ref.[280]. (a) Transient magnetization map as a function of pump–probe delay and depth within the ferrimagnetic layer, showing the formation of two oppositely magnetized regions during reversal. The white region corresponds to zero magnetization and marks the transient boundary separating the switched and unswitched parts of the layer. (b) Normalized magnetization dynamics measured in 1 nm-thick regions near the top, centre, and bottom of the film, evidencing the delayed reversal of deeper regions and the transient coexistence of opposite magnetic states across the thickness.*

Finally alloy engineering with additional rare earths (Tb/Dy/Ho)[281] provides a controlled way to adjust the saturation magnetization $M_s$, uniaxial anisotropy constant $K_u$, and the multidomain threshold, and can substantially widen the compositional window for AO-HIS while reducing the minimum required Gd content (down to ~1.5% reported in RE-mixed alloys). In this framework, the practical criterion becomes the competition between the switching threshold and the multidomain threshold (often summarized as $F_{\text{switch}}(\tau) < F_{\text{multi}}(\tau)$), rather than "being at compensation" per se[281].

### 3.2. Gd based bilayers

With gadolinium still as an essential ingredient, it has also been observed in synthetic ferrimagnets (Gd/Co bilayers)[282,283]. In those systems single-pulse helicity-independent AOS provides a particularly clear bridge between the RE-TM alloys and interface-engineered heterostructures. As illustrated in Fig 16 from Ref.[284], Co/Gd exhibits a highly robust *toggle* behavior: Kerr microscopy shows that the switched domain depends on whether an even or odd number of pump pulses is applied, and the "field-free" TR-MOKE protocol (detecting only every other pulse) resolves the intrinsic ultrafast dynamics without any reset field. The reversal occurs within a few-tens of picoseconds (zero-crossing within a few ps), while the subsequent evolution can include a slower recovery and even a *backswitch* on the ~$10^2$ ps timescale, which becomes strongly field dependent despite the much larger exchange fields governing the initial reversal. This behavior is consistent with a picture where the Co layer dominates the magneto-optical response and is exchange-coupled to a thin, proximity-magnetized Gd region at the interface; AOS is then driven by the strong contrast in ultrafast demagnetization between Co and Gd and by inter-sublattice angular-momentum exchange localized near the interface, closely analogous to the sublattice-driven reversal in amorphous GdCo(GdFeCo) alloys[260].

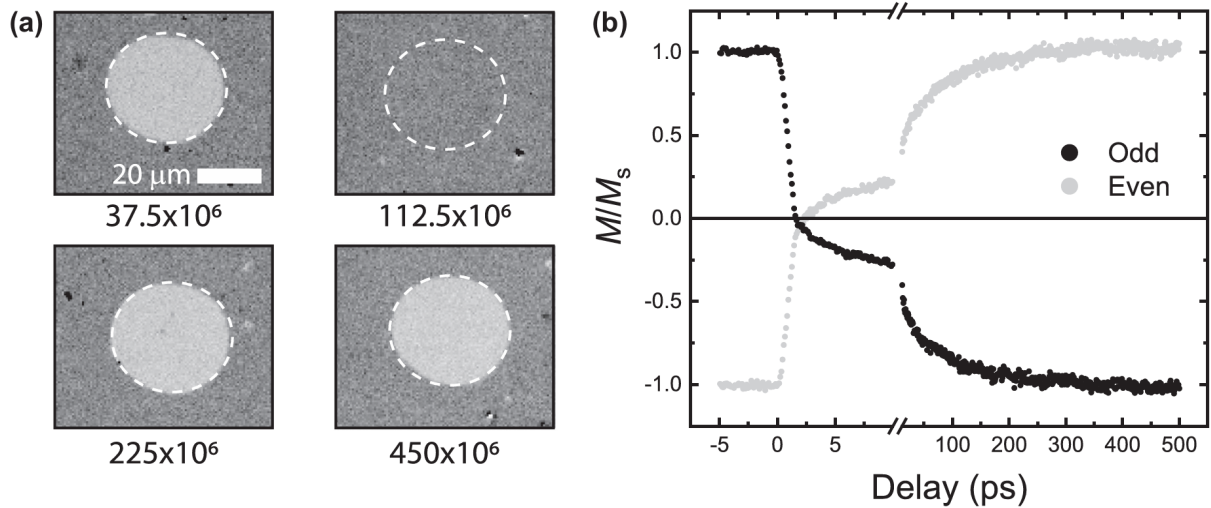

*Figure 16. (from Fig. 1 of Ref.[284]). (a) Wide-field Kerr microscope images of all-optical switching in a Co/Gd bilayer after repeated excitation at 125 kHz. Dark/bright contrast corresponds to opposite out-of-plane magnetization directions. The dashed circle marks the region where the local laser fluence exceeds the switching threshold. A toggled switched domain is observed depending on whether the total number of pump pulses is odd or even (labels indicate the approximate pulse number; scale bar: 20 μm). (b) Field-free time-resolved Kerr measurement of all-optical switching obtained by recording only every second pulse (100 kHz). Selecting odd or even pump pulses isolates the two opposite reversal branches (up→down and down→up, respectively). The magnetization is shown normalized to the saturation value, and an axis break highlights the ultrafast (few-ps) and slower ($10^2$-ps) parts of the switching dynamics.*

Crucially, the link between Co/Gd bilayers and GdCo alloys is not only conceptual but also *materials-realistic*: sputtered Co/Gd interfaces are generically intermixed, producing an alloyed Gd-Co region whose composition and thickness can tune the switching threshold and even the apparent mechanism. This is made explicit in Fig. 17 of Ref.[283], where the interface is modeled as a graded/finite-thickness $Gd_{0.5}Co_{0.5}$ intermixing region (2 or 4 monolayers): compared to an abrupt interface, intermixing reduces the threshold fluence (≈25% for thin Co layers in the model) while preserving the qualitative "switch / no-switch" phase-diagram structure. Within this framework, intermixing enhances the effective Co-Gd angular-momentum transfer at early times and lowers the effective Curie temperature in the intermixed region, making it easier to nucleate reversal at the interface; the subsequent reversal proceeds as a propagating front of reversed Co magnetization through the Co layer, so the bilayer progressively acquires an "alloy-like" character in the active interfacial region. In the limit of extended intermixing, the system approaches a true GdCo alloy locally; conversely, when the intermixed region becomes too close to a homogeneous 50/50 alloy far from compensation, switching can be suppressed—highlighting that what is often labeled "Co/Gd bilayer AOS" is, in practice, strongly governed by the *effective GdCo interfacial alloy* created by interdiffusion.

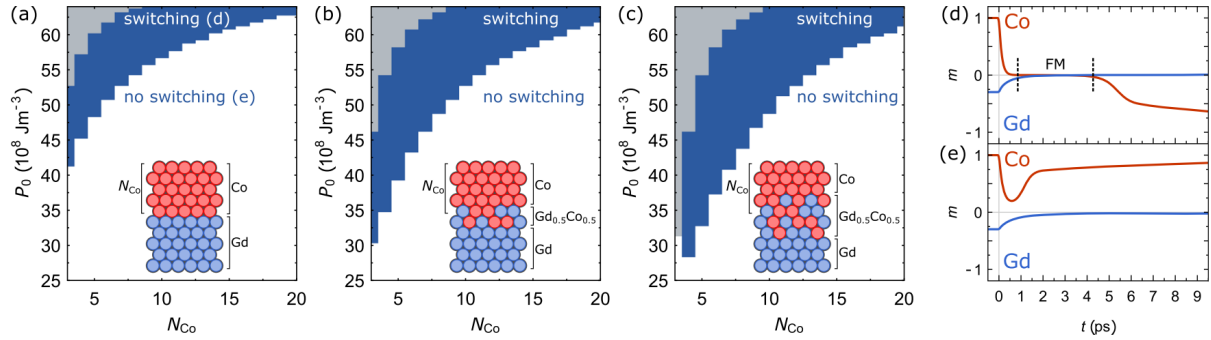

*Figure 17. (from Ref.[283]). Simulated single-pulse all-optical switching (AOS) maps for a Co/Gd synthetic ferrimagnet, plotted versus laser-pulse energy density $P_0$ and the number of Co monolayers $N_{Co}$. (a) Ideal Co/Gd bilayer with an abrupt interface. (b) Co/Gd with intermixing represented by replacing the two Co layers closest to the interface by a $Gd_{0.5}Co_{0.5}$ alloy region. (c) Same as (b) but with a four-layer intermixed region. Blue areas correspond to a switched final state, white areas to no switching, and grey areas indicate conditions where the peak phonon temperature exceeds the Curie temperature. Insets sketch the modeled layer stacks. (d,e) Normalized magnetization dynamics $m(t)$ of the Co (red) and Gd (blue) sublattices for representative points inside the switching region (d) and in the no-switching region (e).*

This helicity-independent all-optical switching (AO-HIS) remains one of the fastest known method for magnetic reversal and one of the most energy-efficient, with bit-switching energies well below 10 fJ. Due to the strong non-equilibrium nature of the reversal as well as the energy absorption required (a so-called "thermal" process), the relevant timescale for magnetization reversal is conveniently set as the minimum time delay between two subsequent laser pulses to achieve two magnetization reversal events. Depending on the materials and exact sample structures, this timescale varies from a few ps to several hundreds of ps[123,285–287]. Note that a similar effect has been observed more than a decade later in the ferrimagnetic half metal Heusler alloy $Mn_2Ru_xGa$[122,124].

### 3.3. RE-TM heterostructure

### 3.3.1 All optical Helicity independent switching

A major recent development is that Gd-based materials are not the only platform exhibiting deterministic single-shot switching. By mixing Gd with heavier rare earths (Dy, Tb, Ho), AO-HIS can be preserved while tuning magnetic parameters (anisotropy, exchange, saturation magnetization) and even reducing the required Gd content to very small fractions, thereby widening the materials space for device-relevant stacks[281]. In parallel, substitution studies show that when Gd is progressively removed, switching becomes incomplete or disappears in several RE-TM alloys (TbCo and DyCo in Ref.[288]), highlighting that angular-momentum availability and transfer pathways, determine whether a deterministic toggle reversal is obtained.

A particularly instructive Gd-free example is provided by Co/Ho multilayers, where Peng *et al.*[289] demonstrated single-shot AO-HIS in wedge-shaped $[Co/Ho]_N$ stacks with PMA, and established a pulse-duration-fluence state diagram that is surprisingly close to the Gd-based case despite Ho's stronger spin-orbit coupling. In Figure 18 experiments show robust toggle switching for a given Co and Ho thickness. It is true in a restricted thickness window, while thicker regions tend toward irregular domains and eventual multidomain formation under repeated excitation. This result expands AO-HIS to an additional RE-TM architecture and suggests that the "triangular" window is a robust emergent feature whenever reversal competes with multidomain breakup.

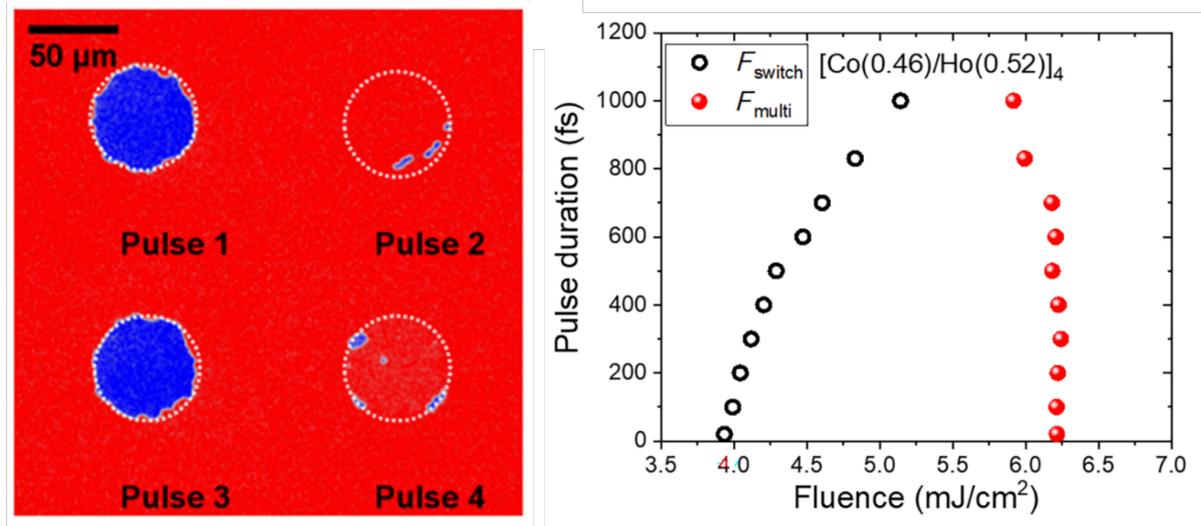

*Figure 18. Single-shot AO-HIS in Co/Ho multilayers. Left: Kerr microscopy images after successive single-pulse excitations in [Co(0.46)/Ho(0.52)]$_4$ (fluence 5.22 mJ·cm$^{-2}$, pulse duration 50 fs), where each pulse writes a switched spot (toggle behavior). Red/blue contrast indicates opposite out-of-plane magnetization directions. Right: pulse-duration-fluence state diagram showing the switching threshold $F_{switch}$ (open symbols) and multidomain threshold $F_{multi}$ (filled symbols), highlighting a triangular operational window for deterministic switching. (From Fig. 1 and Fig. 5 of Ref.[289]).*

At the same time, the *dynamics* of AO-HIS can change qualitatively in Gd-free alloys. Kunyangyuen *et al.*[290] reported deterministic single-pulse toggle switching in CoHo and CoDy single-layer alloys, but demonstrated that the reversal mechanism is fundamentally different from the ultrafast (ps) exchange-driven pathway of GdFeCo: time-resolved measurements reveal only ultrafast demagnetization and recovery within the sub-ns window, while the actual magnetization reversal develops on microsecond timescales through domain-wall reorganization and domain coalescence, seeded by the femtosecond excitation and evolving long after the pulse.

A unifying framework emerges from angular-momentum reservoir engineering in simple bilayers. In Co/Gd, Kunyangyuen *et al.*[291] showed that the switching speed can be tuned over more than three orders of magnitude by varying the Gd thickness (reservoir size) as shown in Figure 19 or inserting an interfacial Pt spacer (transfer bottleneck). Increasing the Gd thickness yields few-ps reversal consistent with efficient angular-momentum transfer during Gd demagnetization, whereas thinning Gd or inserting Pt suppresses transfer and drives the system toward slow, domain-growth-mediated reversal on ns–μs timescales—continuous with the behavior observed in CoDy and CoHo alloys.

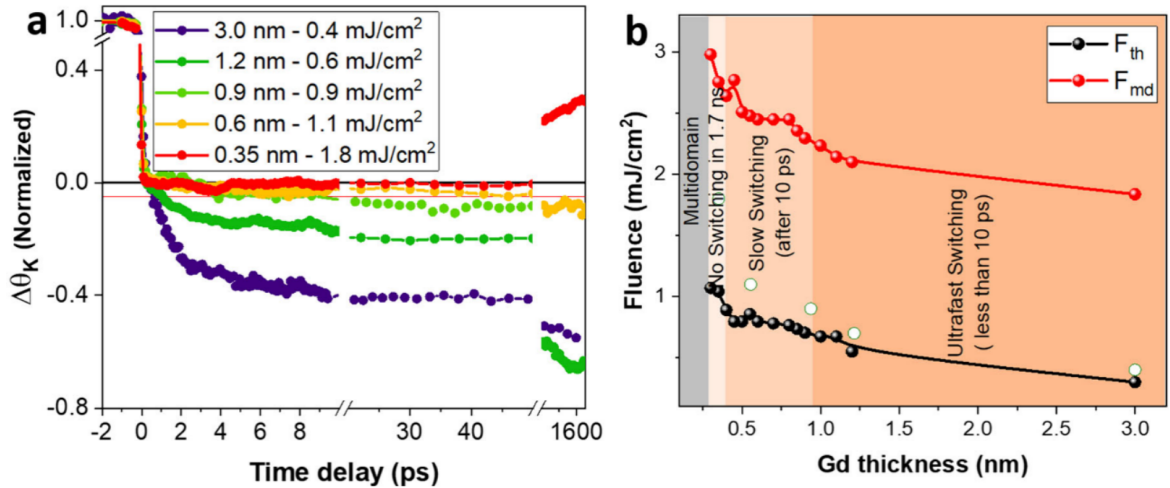

*Figure 19.* Influence of Gd thickness on reversal dynamics and switching thresholds in Co(0.7 nm)/Gd($t_{Gd}$)/Pt. (a) Time-resolved magneto-optical Kerr effect (TR-MOKE) traces measured on Si/SiO2(500)/Ta(3)/Pt(3.7)/Co(0.7)/Gd($t_{Gd}$)/Pt(3) wedge samples, excited by a single 50 fs laser pulse under an out-of-plane applied field of 60 mT. The dynamics evolve from ultrafast reversal for thick Gd (ps-scale zero-crossing and switching) to incomplete/slow reversal for thinner Gd, with a long-lived demagnetized or partially reversed state. (b) Threshold fluence for AO-HIS switching, $F_{th}$ (black), and for multidomain formation, $F_{md}$ (red), as a function of Gd thickness $t_{Gd}$ for 50 fs pulses. Background shading indicates the dynamic regime: no switching within the experimental time window (delay line too short), slow switching (crossing 5% switching after 10 ps), and ultrafast switching (crossing 5% switching within 10 ps). (From Fig. 2 of Ref.[291])

These results directly support the central message that AO-HIS is governed by (i) how much angular momentum is available in the RE subsystem and (ii) how efficiently it can be transferred to the TM during the nonequilibrium demagnetization window. In this picture, Gd remains uniquely efficient (weak SOC, favorable demagnetization/transfer balance), while heavier REs can still enable deterministic reversal but typically with reduced transfer efficiency and therefore slower, domain-mediated pathways unless the stack is engineered to compensate.

### 3.3.2 All optical Precessional switching

An additional switching modality has recently been identified in several RE-TM multilayers and alloys, where a single laser pulse triggers magnetization reversal through a laser-induced spin reorientation followed by a precessional trajectory[116], rather than through the ultrafast exchange-driven "toggle" pathway established in GdFeCo. In these Tb- (and Dy-) based systems, the decisive ingredient is a fast, transient reduction (and/or tilting) of the effective perpendicular anisotropy after optical heating, which redirects the effective field toward the film plane and launches a large-angle precession that carries the magnetization across the energy barrier on tens to hundreds of picoseconds. This picture naturally explains two striking experimental hallmarks: (i) the appearance of concentric ring ("bullseye") domain patterns at higher fluence, consistent with a fluence-dependent precessional response in the presence of a strong spatial temperature gradient, and (ii) the weak dependence of the switching thresholds on pulse duration over a broad range, indicating that the final state is governed primarily by the energy transferred to (and stored in) the lattice/anisotropy landscape, rather than by the peak electronic temperature.

These features are clearly illustrated in Fig. 20 (from Ref.[292]), where single-pulse switching is obtained in a representative [Tb/Fe] multilayer: at low fluence a single-domain toggle is observed, while increasing fluence generates additional reversed rings and ultimately a multidomain center; the associated state diagram shows that the main fluence thresholds delimiting these regimes remain nearly unchanged from 50 fs up to ~12 ps, supporting a precessional mechanism driven by anisotropy quenching/reorientation rather than by a purely non-thermal ultrafast sublattice exchange scenario. Very recently, this precessional route has also been extended to ferromagnetic trilayers. Xu et al.[293] demonstrated single-shot optical precessional switching in Pt/Co/Pt under an in-plane magnetic field, and showed that the reversal is governed by thermal anisotropy torque, with the switching window strongly influenced by PMA and thermal relaxation.

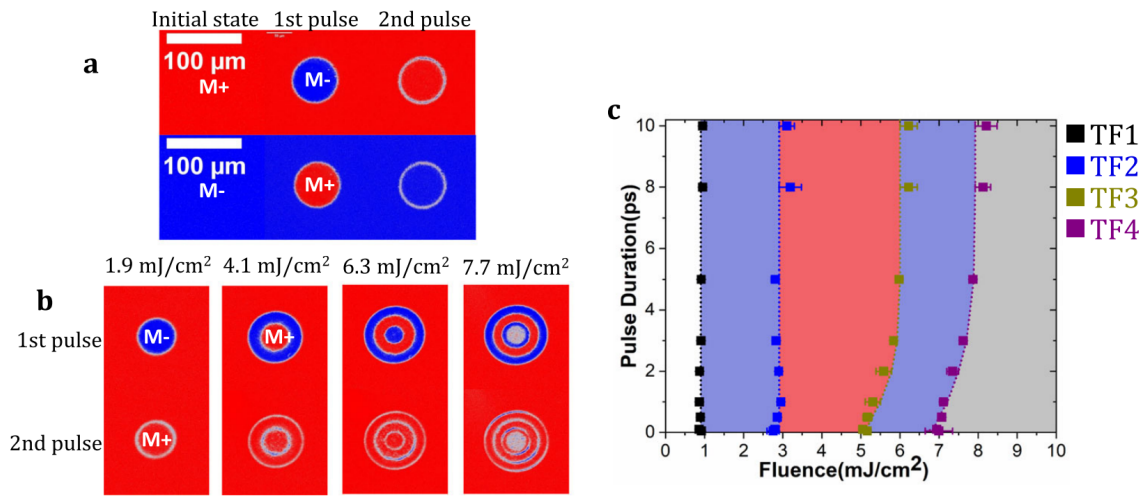

*Figure 20.* Single-pulse switching and pulse-duration/fluence state diagram evidencing a precessional pathway in a Tb-based multilayer (Figure reproduced from Fig. 1 of Ref.[292]). (a) Background-subtracted Kerr microscopy images after a single 50-fs pulse (fluence 1.9 mJ/cm²), starting from opposite initial magnetization states, showing deterministic reversal of the central domain. (b) Images after a first single 50-fs pulse at increasing fluence, revealing the emergence of concentric ring ("bullseye") magnetic textures and a progression from single-domain reversal to multi-ring and ultimately multidomain states. (c) State diagram of the final magnetic configuration as a function of pulse duration and fluence, defining thresholds (TF1–TF4) for switching and successive ring formation; the weak dependence of these thresholds on pulse duration supports a lattice/anisotropy-driven in-plane reorientation followed by precessional reversal rather than an ultrafast exchange-driven toggle mechanism.

In Ref.[294], field-free single-shot reversal is demonstrated in [Tb/Co]-based p-MTJs, and a macrospin model coupled to a two-temperature description reproduces the main trends by combining (i) a rapid temperature-driven drop of the uniaxial anisotropy and (ii) a small tilt of the anisotropy axis, which together produce an in-plane reorientation and subsequent precession toward the reversed state. Fig. 21 summarizes this modeling framework and its comparison to time-resolved MOKE images, emphasizing that the crucial reversal dynamics develop on the longer (ps–100 ps) timescale where anisotropy-driven precession dominates, consistent with the experimentally observed pulse-duration robustness.

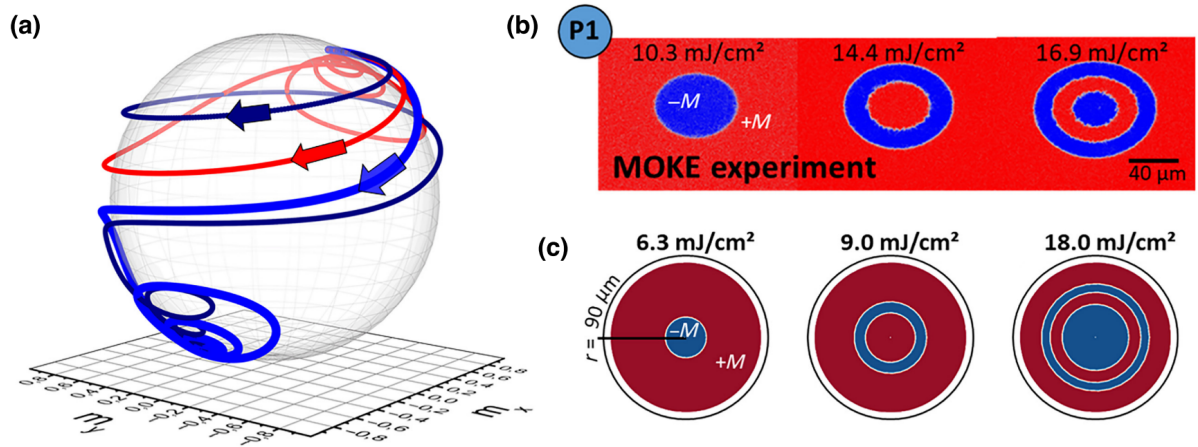

*Figure 21.* Macrospin description of in-plane reorientation-driven precessional switching in Tb/Co-based structures (Figure reproduced from Fig. 2a–c of Ref.[294]). (a) Three-dimensional view of the simulated magnetization trajectory during the laser-driven reversal predicted by a Landau-Lifshitz-Gilbert macrospin model coupled to a two-temperature description, highlighting the role of a rapid, temperature-induced reduction of the uniaxial anisotropy (and a slight anisotropy-axis tilt) in launching a large-angle precession. (b) Experimental background-subtracted time-resolved MOKE images at a representative position (P1) for a 50-fs excitation at increasing fluence, showing the evolution of the magnetization contrast consistent with a reorientation/precessional scenario. (c) Simulated magnetization pattern at 100 fs for a single Gaussian laser spot assuming fully independent spins, illustrating the early-time spatial profile that seeds the subsequent anisotropy-driven dynamics.

Finally, time-resolved single-shot studies further show that Tb/Co multilayers stabilize the switched configuration on ~50–100 ps timescales and exhibit fluence- and composition-dependent dynamics that do not map onto the standard Gd-based single-shot framework, reinforcing the view that Tb-based single-pulse switching can access a distinct, precession-mediated route whose efficiency depends sensitively on anisotropy engineering and angular-momentum dynamics in RE environments[295].

Having established different pathways to efficiently reverse magnetization of single magnetic materials with a single pulse, the next step is to integrate this physics in actual devices. Two key aspects are then to (i) trigger and (ii) probe the switching electronically without the direct need for an optical femtosecond pulse. We dive into this in the next section.

## 4. Ultrafast Spintronics: From Femtomagnetism to Functional Devices

The realization that femtosecond excitation can not only quench magnetization but also trigger deterministic magnetization reversal[131,135], generate ultrafast spin currents[35,81–83,176–178], and reconfigure magnetic order across complex heterostructures has established ultrafast spintronics as a natural extension of femtomagnetism toward device-oriented functionality. In this emerging paradigm, optical or hot-electron excitation drives magnetic systems far from equilibrium, enabling angular-momentum transfer and spin transport on timescales well below those accessible to conventional spin-torque mechanisms.

Within this framework, two main classes of ultrafast spintronic devices can be distinguished.
(i) In the first class, the active magnetic layer itself—typically a ferrimagnetic rare-earth-transition-metal alloy such as Gd-based systems or Tb/Co multilayers—undergoes direct all-optical switching following femtosecond excitation.
(ii) In the second class, ultrafast demagnetization of an optically excited layer is exploited as a source of angular momentum that is subsequently transferred to an adjacent ferromagnetic layer via ultrafast spin currents, enabling indirect but deterministic switching.

In the latter case, a central consequence of ultrafast demagnetization is the generation of a femtosecond burst of angular momentum that gives rise to ultrafast spin pumping and superdiffusive spin transport. Whether initiated by direct optical absorption or by hot-electron injection, the rapid deposition of energy leads to a transient collapse of magnetic order, which in turn launches spin-polarized currents capable of propagating through metallic spacers, insulating layers, or tunnel barriers on sub-picosecond timescales[72,133,178,232,240,296]. These ultrashort spin-current pulses act as an ultrafast analogue of spin-transfer torque, but operate three orders of magnitude faster and without the need for sustained charge currents.

Building on these principles, a series of spintronic device concepts has been developed to harness all-optical switching (AOS) and ultrafast spin injection in functional architectures, including spin valves, racetrack-type structures, and magnetic tunnel junctions. In many of these designs, the ultrafast switching of one magnetic layer—typically a ferrimagnetic or exchange-coupled layer—serves as the source of angular momentum that deterministically drives the reversal of an adjacent ferromagnetic layer. Such schemes enable magnetic switching at repetition rates approaching the terahertz limit while operating at energies far below those required for conventional spin-transfer-torque or spin-orbit-torque devices.

## 4.1 Optical switching of one layer in MTJ or racetrack memory devices

In recent years, strong interest has emerged in integrating all-optical switching (AOS) into spintronic device architectures, motivated by the prospect of ultrafast, field-free, and energy-efficient magnetic writing[297,298]. One prominent approach consists in the development of AOS-compatible magnetic tunnel junctions (AOS-MTJs), in which the free layer can be deterministically switched by a single femtosecond laser pulse while preserving reliable electrical readout through tunneling magnetoresistance (TMR). Early demonstrations based on amorphous GdFeCo free layers established the feasibility of optically switchable MTJs and direct photonic-to-magnetic information conversion as shown in Figure 22, but suffered from low TMR ratios, weak perpendicular magnetic anisotropy, and limited scalability[299].

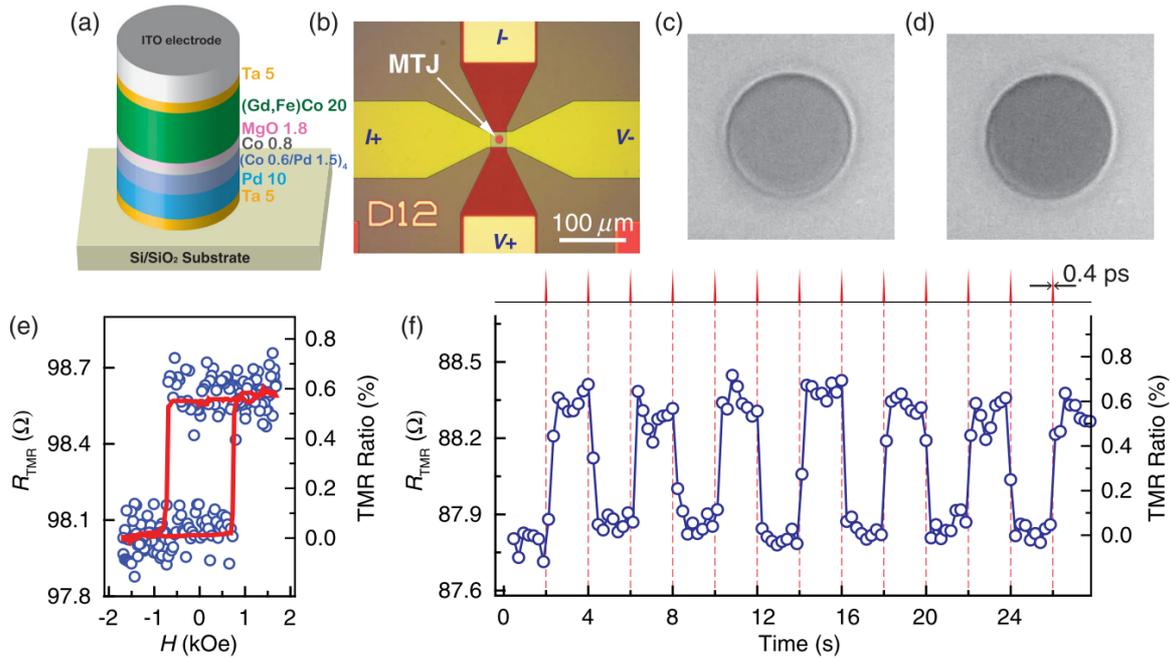

*Figure 22. All-optical switching of a magnetic tunnel junction using subpicosecond laser pulses. (a) Schematic of the AOS-compatible MTJ stack incorporating a Gd(Fe,Co) free layer and an MgO tunnel barrier. (b) Optical microscope image of a representative MTJ device with a transparent ITO top electrode enabling optical access. (c,d) Magneto-optical Kerr effect (MOKE) images of the MTJ pillar before and after excitation by a single 0.4-ps laser pulse, showing complete reversal of the Gd(Fe,Co) free layer. (e) Minor tunneling magnetoresistance $R_{TMR}(H)$ loop measured under a perpendicular magnetic field. (f) Time-resolved $R_{TMR}$ during exposure to single laser pulses (0.5 Hz), demonstrating deterministic optical switching between high- and low-resistance states. Adapted from Chen et al.[299].*

More recent advances have shown that synthetic ferrimagnet-based free layers, such as Co/Gd bilayers[300] and [Co/Tb]$_n$ multilayers[113,294], constitute a far more favorable materials platform. In these systems, the optically switchable ferrimagnetic layer is exchange-coupled to a CoFeB/MgO interface, thereby combining robust single-pulse helicity-independent AOS with the high spin polarization and interfacial anisotropy required for large TMR values. This strategy has enabled TMR ratios exceeding 30–70%, sub-10-ps switching times, and energy costs in the sub-100 fJ/bit range, while remaining compatible with nanofabrication and CMOS-oriented MTJ stacks[294,300].

Another representative approach consists in interfacing all-optical switching with spintronic racetrack memories, enabling direct conversion of photonic signals into magnetic information without intermediate electronics[301–303]. In this concept, femtosecond laser pulses are used to locally write magnetic domains via single-pulse AOS in synthetic ferrimagnetic racetracks, while subsequent domain-wall motion driven by spin-orbit torques enables information transport along the nanowire. Lalieu et al. demonstrated deterministic single-pulse AOS in Pt/Co/Gd racetracks combined with efficient spin Hall effect-driven domain-wall motion, allowing on-the-fly optical writing and electrical transport of magnetic bits[301]. Building on this idea, Pezeshki et al. proposed and numerically demonstrated hybrid plasmonic-photonic architectures that focus guided light beyond the diffraction limit, enabling localized AOS and magneto-optical readout of nanoscale domains within racetrack geometries integrated on photonic waveguides[302,303]. Very recently, experimental progress has been reported toward monolithic integration of AOS within photonic integrated circuits, where guided femtosecond pulses induce deterministic switching of Co/Gd microstructures patterned directly on top of dielectric waveguides[304].

Beyond local optical switching of magnetic layers, helicity-dependent femtosecond excitation can also be used to control domain-wall motion in ferromagnetic wires. In Co/Ni/Co-based systems[305,306], synchronized laser and current pulses were shown to enable helicity-selective domain-wall depinning and propagation through the combined action of optical effects and spin torques. These hybrid optical-electrical schemes reduce the current threshold for domain-wall motion and can be further extended to optoelectronic logic-in-memory concepts, including the demonstration of Boolean functions such as AND, OR, NAND, and NOR[307].

Together, these works establish racetrack memories as a promising platform for hybrid photonic-spintronic devices, combining ultrafast optical writing with current-driven domain-wall motion for scalable, high-speed information processing.

### 4.2 Hot-electron-driven demagnetization and magnetization reversal

Ballistic hot electrons generated by femtosecond optical excitation can induce ultrafast demagnetization in Pt/Cu/(Co/Pt) multilayers[162]. As illustrated in Fig. 23 and in the corresponding schematic, photoexcited carriers created in the Pt layer by light absorption propagate ballistically through the Cu spacer over distances up to ~300 nm within a few hundred femtoseconds. These hot electrons efficiently transfer heat to the buried ferromagnetic (Co/Pt) multilayer, leading to demagnetization on a sub-picosecond timescale. Remarkably, the resulting demagnetization dynamics is comparable in speed and efficiency to that obtained under direct optical excitation of the ferromagnetic layer itself, demonstrating that hot-electron transport alone can act as an ultrafast and efficient driver of magnetization dynamics[162].

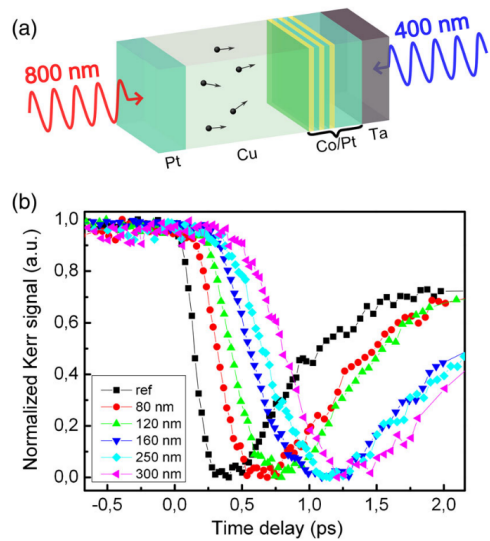

*Figure 23.* (a) Schematic of the Pt/Cu/(Co/Pt) multilayer structure used to investigate hot-electron-induced ultrafast demagnetization. Femtosecond laser pulses (800 nm) are absorbed predominantly in the top Pt layer, generating hot electrons that propagate ballistically through the Cu spacer toward the buried (Co/Pt) multilayer. (b) Time-resolved magneto-optical Kerr effect (TR-MOKE) measurements of the normalized Kerr signal for different Cu thicknesses (0–300 nm). A clear demagnetization is observed even for the thickest Cu spacers, where direct optical excitation of the ferromagnetic layer is negligible. The increasing delay and broadening of the demagnetization onset with Cu thickness provide direct evidence of ballistic hot-electron transport driving ultrafast demagnetization on a sub-picosecond timescale. Adapted from Ref.[162]

In a closely related architecture, hot-electron excitation can even induce deterministic magnetization reversal in buried magnetic layers, closely mirroring the dynamics of single-pulse all-optical switching[135,308,309]. In Pt/Cu/GdFeCo heterostructures, femtosecond optical excitation of the Pt/Cu layers generates a hot-electron pulse that propagates ballistically through the Cu spacer over distances of several tens of nanometres within a few hundred femtoseconds. These carriers trigger ultrafast demagnetization and subsequent deterministic reversal of the buried GdFeCo ferrimagnet on the same characteristic timescale of approximately 5 ps as observed under direct optical excitation[135]. This demonstrates that energetic carriers, rather than photons, can transport sufficient energy and angular momentum to drive the demagnetization and reversal process required for GdFeCo switching, as illustrated in Fig. 24, adapted from Ref. [135].

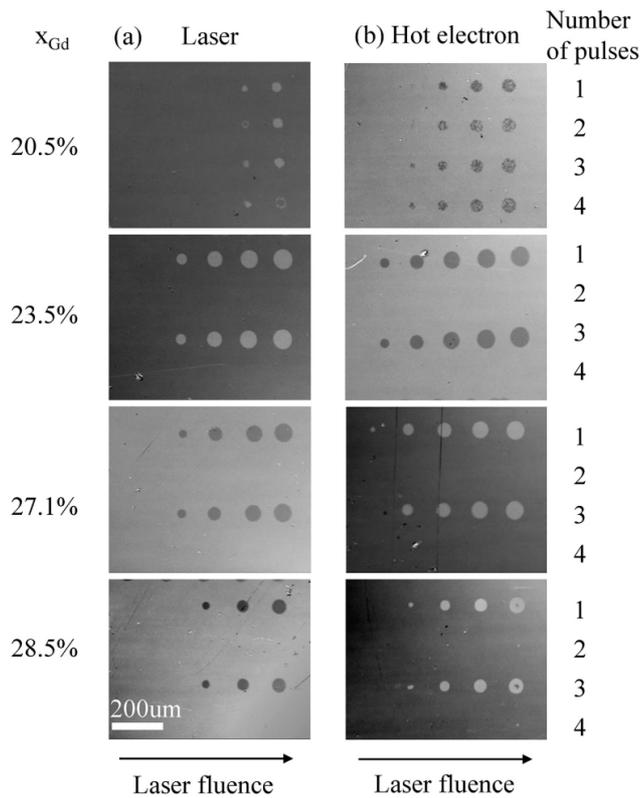

*Figure 24.* Magneto-optical Kerr microscopy images of single-pulse magnetization switching in $Gd_x(FeCo)_{1-x}$ ferrimagnetic films induced by (a) direct femtosecond laser excitation and (b) femtosecond hot-electron excitation. Images are shown after one, two, three, and four successive pulses for several Gd concentrations above and below the compensation composition. In both excitation schemes, deterministic single-pulse switching is observed for compositions close to compensation, while multi-domain states appear for compositions farther from compensation. The qualitative similarity between light-induced and hot-electron-induced switching demonstrates that ballistic hot electrons can efficiently trigger the same ultrafast reversal mechanism as direct optical excitation. Adapted from Ref.[135].

A particularly important result further clarifies the thermal origin of helicity-independent switching. Yang *et al.*[309] demonstrated that magnetization reversal in GdFeCo can also be triggered by picosecond electrical current pulses, without any optical excitation. In their experiment, a photoconductive switch was used to generate electrical pulses of approximately 9–10 ps duration, which produced ultrafast Joule heating in a patterned GdFeCo device. Remarkably, a single electrical pulse was sufficient to induce deterministic magnetization reversal, with dynamics comparable to those observed in laser-driven experiments. The temporal profile of the electrical pulse, shown in Fig. 25, demonstrates that the magnetization switching occurs even when the excitation pulse is two orders of magnitude longer than the femtosecond optical pulses typically used in all-optical switching experiments.

This result has two important implications. First, it confirms that ultrafast switching in RE-TM ferrimagnets is fundamentally a thermal process, driven by rapid energy deposition in the electronic system rather than by any specific optical mechanism. Second, it is highly significant from a technological perspective: generating picosecond electrical pulses is considerably easier and more compatible with integrated electronics than producing femtosecond optical pulses. Electrical excitation therefore provides a practical route toward device implementations of ultrafast thermally driven magnetic switching.

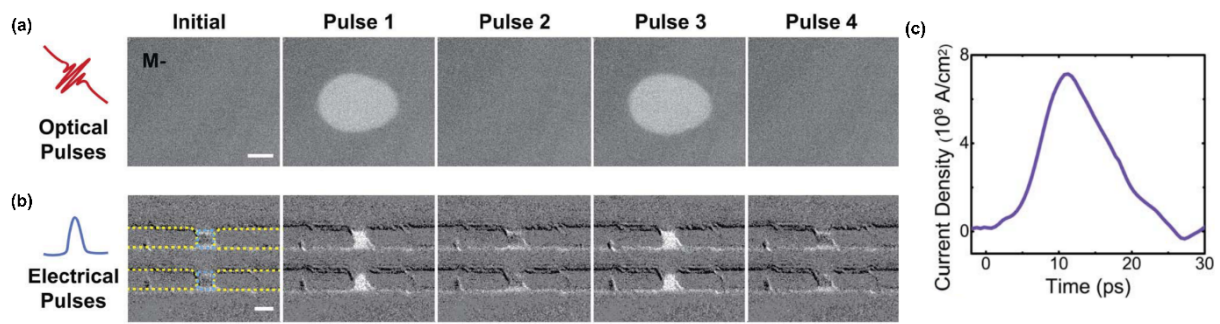

*Figure 25.* Ultrafast magnetization reversal induced by optical pulses (a) and picosecond electrical pulses (b) in GdFeCo, adapted from Yang et al.[309]. In (b), a photoconductive switch generates ~9-ps electrical pulses that propagate along a coplanar stripline and produce ultrafast Joule heating in the GdFeCo section. Despite the relatively long pulse duration compared with femtosecond optical excitation, a single electrical pulse, shown in (c), is sufficient to trigger deterministic magnetization reversal, demonstrating that the switching mechanism is primarily thermally driven.

### 4.3 Ultrafast spin-transfer in spinvalve and junction

Femtosecond demagnetization acts as a powerful source of spin-polarized carriers, launching ultrafast spin currents capable of inducing magnetization reversal in adjacent magnetic layers. In spin-valve heterostructures, a single femtosecond laser pulse can demagnetize one magnetic layer and generate a spin current that propagates through non-magnetic spacers over distances reaching 50–80 nm, leading to the reversal of a second magnetic layer within a few hundred femtoseconds[130,131,310].

The first clear evidence that all-optical switching (AOS) can be mediated by ultrafast spin transfer in a spin-valve geometry was obtained in GdFeCo/Cu/(Co/Pt) heterostructures[127]. In these systems, femtosecond laser excitation of the ferrimagnetic GdFeCo layer produces an ultrafast demagnetization that generates spin-polarized hot-electron currents, which propagate non-locally across the metallic Cu spacer and inject angular momentum into the adjacent ferromagnetic Co/Pt multilayer. As a result, the magnetization of the Co/Pt layer can be reversed without direct optical switching of the ferromagnet itself, demonstrating a spin-current-mediated, non-local AOS mechanism.

A key outcome of this work is that the two magnetic layers can be addressed selectively using a *single* femtosecond laser pulse, solely by adjusting the excitation fluence. As shown in Fig. 26, moderate-fluence pulses induce helicity-independent single-pulse switching of the GdFeCo layer alone, leaving the Co/Pt layer unchanged, whereas higher-fluence pulses lead to the simultaneous reversal of both GdFeCo and Co/Pt. This enables controlled access to the four remanent magnetic configurations of the spin valve (P⁺, P⁻, AP⁺, AP⁻) through a sequence of single optical pulses under zero applied magnetic field. Systematic control experiments, including variation of the Cu spacer thickness and insertion of Pt spin-scattering layers, further confirmed that the switching of the Co/Pt layer is governed by spin-polarized currents generated in GdFeCo rather than by static exchange or dipolar coupling. Inserting a VO$_2$ layer in the Cu spacer can further allow or block the magnetization reversal by tuning the ambient temperature respectively above or below the metal-insulator phase transition temperature[311]. These results establish ultrafast spin-current generation during laser-induced demagnetization as the physical mechanism enabling multi-level, non-local all-optical switching in spin-valve structures.

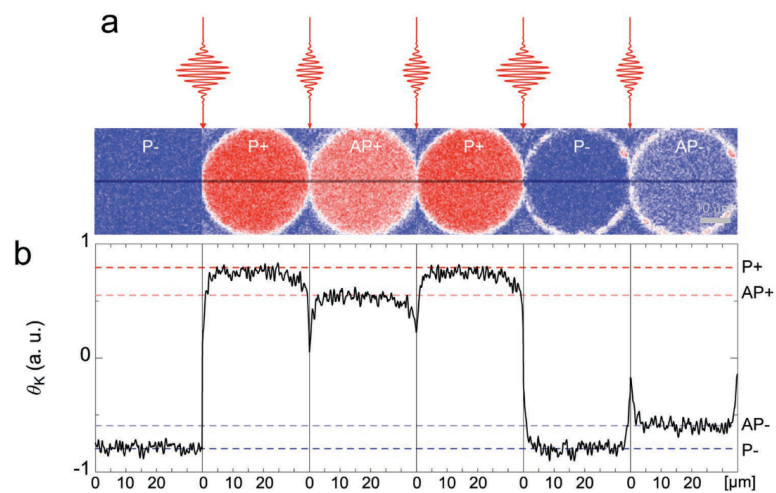

*Figure 26. Single-pulse multi-level all-optical switching in a GdFeCo/Cu/[Co/Pt] spin-valve structure. Figure reproduced from Fig. 2 of Ref.[127]. A sequence of single femtosecond laser pulses with different fluences enables selective and deterministic access to the four remanent magnetic configurations of the spin valve (P⁺, AP⁺, AP⁻, P⁻). Depending on the pulse energy, a single optical excitation can switch either the ferrimagnetic GdFeCo layer alone or both the GdFeCo and ferromagnetic [Co/Pt] layers. The switching of the [Co/Pt] layer is mediated by an ultrafast spin current generated during the laser-induced demagnetization of GdFeCo and transmitted through the Cu spacer, demonstrating non-local, spin-transfer-driven all-optical switching in a spin-valve geometry.*

Subsequent time-resolved studies revealed that spin-current-driven switching of a ferromagnetic layer can occur on remarkably short timescales, well below one picosecond. In particular, Remy *et al.*[130] demonstrated sub-picosecond magnetization reversal of a Co/Pt multilayer driven by a non-local spin current generated in an adjacent GdFeCo layer. In this work, the ultrafast demagnetization of GdFeCo produces a bipolar spin-polarized hot-electron current that injects angular momentum into the ferromagnetic Co/Pt layer, providing an external source of spin angular momentum precisely when the magnetization amplitude approaches zero.

As shown in Fig. 27 time-resolved magneto-optical Kerr effect measurements reveal a complete reversal (zero-crossing) of the Co/Pt magnetization within approximately 400 fs, followed by a rapid recovery toward the reversed state within a few picoseconds. This behavior directly demonstrates that the injected spin current lifts the degeneracy of the paramagnetic state and overcomes the universal critical slowing down that normally limits ultrafast remagnetization near the Curie temperature. The observed dynamics are explained by a combination of ultrafast spin heating prior to reversal and ultrafast spin cooling after reversal, which together enable deterministic and accelerated magnetization switching. These results establish non-local ultrafast spin transfer as an efficient mechanism to drive ferromagnetic magnetization reversal on sub-picosecond timescales, well beyond the limits imposed by purely thermal processes

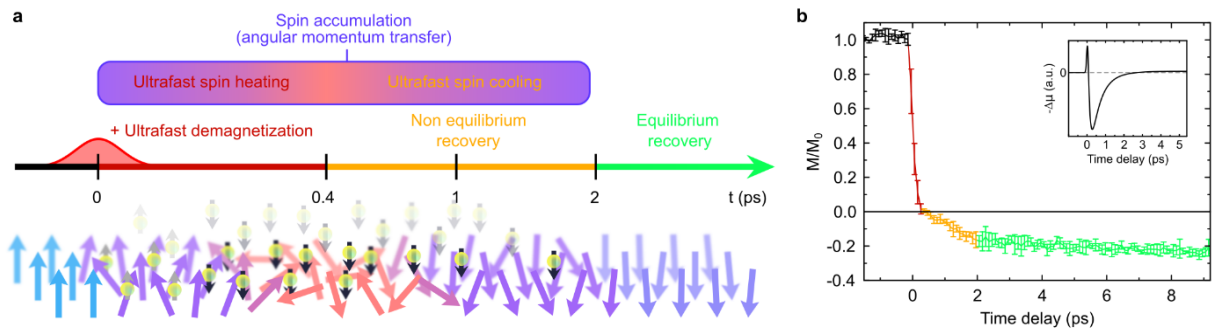

*Figure 27.* Sub-picosecond spin-current-driven magnetization reversal of a ferromagnetic Co/Pt multilayer. Figure reproduced from Fig. 2 of Remy et al.[130]. (a) Schematic illustration of the ultrafast magnetization reversal mechanism of the free Co/Pt layer driven by a non-local spin current generated in an adjacent GdFeCo layer upon femtosecond laser excitation. The ultrafast demagnetization of GdFeCo produces a bipolar spin-polarized current that injects angular momentum into the Co/Pt multilayer, inducing ultrafast spin heating and spin cooling processes. (b) Time-resolved magneto-optical Kerr effect measurements showing the temporal evolution of the Co/Pt magnetization in the parallel (P) configuration. Magnetization reversal (zero-crossing) occurs within ~400 fs, followed by a rapid recovery toward the reversed state on a few-picosecond timescale. These results demonstrate that non-local ultrafast spin transfer overcomes the critical slowing down associated with remagnetization and enables sub-picosecond ferromagnetic switching.

Building on these findings, ultrafast spin-transfer-driven switching was subsequently demonstrated in *purely ferromagnetic* spin-valve structures, establishing that ferrimagnetic layers are not a prerequisite for single-pulse ultrafast reversal. In [Pt/Co]/Cu/[Co/Pt] spin valves, Igarashi *et al.*[131] showed that a single femtosecond laser pulse can induce deterministic magnetization switching of the free Co/Pt layer on sub-picosecond timescales. As illustrated in Fig. 28, both parallel-to-antiparallel (P→AP) and antiparallel-to-parallel (AP→P) switching can be achieved depending on the laser fluence, with a clear asymmetry between the two processes.

At low fluence, single-pulse P→AP switching is systematically observed, even for Cu spacer thicknesses as large as 80 nm, demonstrating the long-range nature of the ultrafast spin current responsible for the reversal. Remarkably, this switching occurs despite the absence of direct optical excitation of the reference layer, ruling out static exchange or dipolar coupling. Time-resolved magneto-optical measurements further reveal that the reversal of the free layer takes place within less than one picosecond, placing it in the same ultrafast regime as spin-transfer-torque-like dynamics but far beyond the speed limits of conventional current-driven STT.

In contrast, AP→P switching is observed only at higher fluences and for thinner Cu spacers, indicating a distinct microscopic origin. The combined static and dynamic measurements support a scenario in which P→AP switching is driven predominantly by ultrafast spin currents generated during the demagnetization of the free layer itself, while AP→P switching involves spin currents emitted during the demagnetization—and partially the remagnetization—of the reference layer. These results demonstrate that ferromagnetic spin valves can host bipolar, ultrafast, single-pulse optical switching mediated by transient spin-current injection, thereby directly bridging the concepts of ultrafast magnetism and spin-transfer torque in technologically relevant spin-valve architectures.

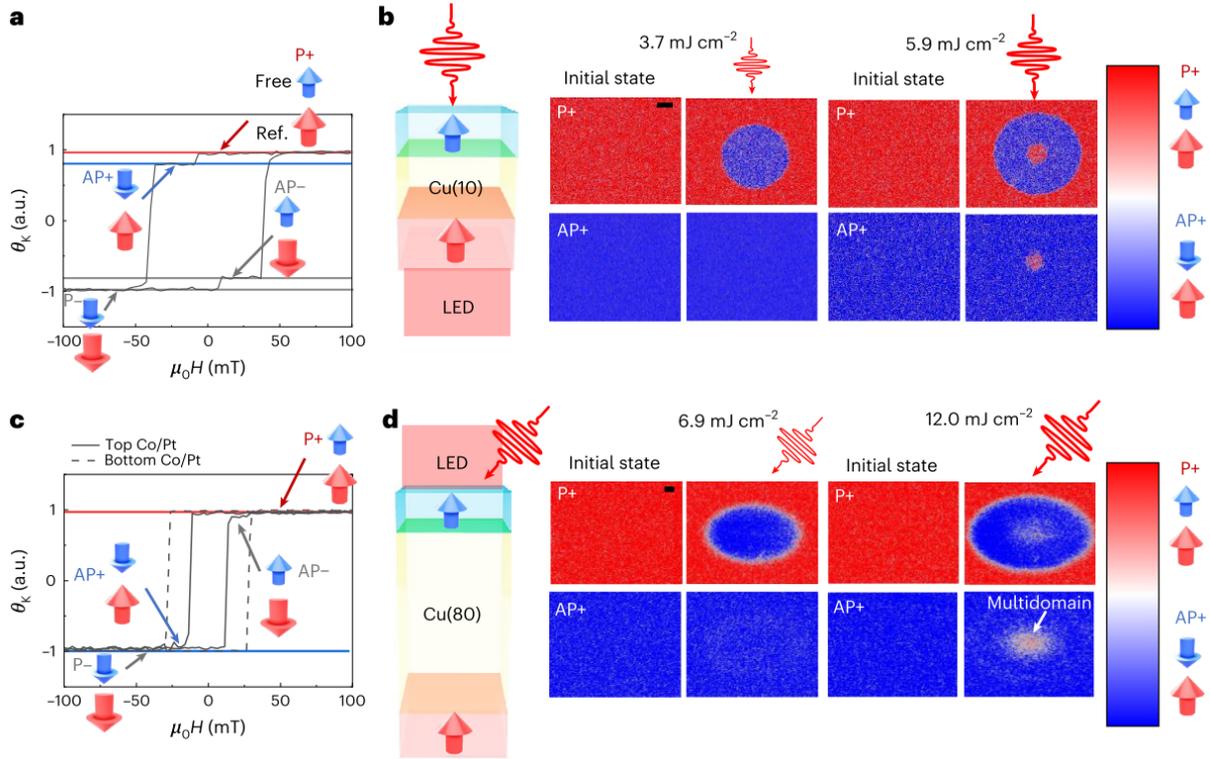

*Figure 28. Single-pulse ultrafast magnetization switching in a ferromagnetic spin-valve structure. Figure reproduced from Fig. 2 of Ref.[131]. (a) Magneto-optical Kerr effect (MOKE) hysteresis loop measured from the reference layer of a [Pt/Co]/Cu/[Co/Pt] spin valve (Cu thickness 10 nm), evidencing four stable magnetic configurations ($P^+$, $AP^+$, $AP^-$, $P^-$). (b) MOKE microscopy images obtained after excitation with a single linearly polarized femtosecond laser pulse, starting from $P^+$ and $AP^+$ initial states. At low fluence, deterministic parallel-to-antiparallel (P→AP) switching of the free Co/Pt layer is observed, whereas at higher fluence antiparallel-to-parallel (AP→P) switching occurs. (c,d) Corresponding measurements for a spin valve with a thick Cu spacer (80 nm), demonstrating that P→AP switching persists even when the reference layer is not optically excited. These results establish that single-pulse, bipolar magnetization switching in ferromagnetic spin valves is mediated by ultrafast spin-current injection over distances up to tens of nanometres.*

A unified microscopic picture of bipolar ultrafast switching in spin valves was later established by disentangling the respective roles of demagnetization- and remagnetization-driven spin transport, in the case of thick non-metallic spacers[312]. By combining time-resolved magneto-optical Kerr rotation and ellipticity measurements with spin-diffusion and extended *s-d* model simulations[132], Singh *et al.*[312] demonstrated that the ultrafast laser excitation of a spin valve generates a time-dependent spin accumulation throughout the multilayer, which directly governs the switching polarity of the free layer. As shown in Fig. 29, the magnetization dynamics is vastly different for the free layer (which reverses or not depending on the orientation of the reference layer) and the reference layer (which only demagnetizes and remagnetizes). The observed difference between the dynamics extracted from Kerr rotation or ellipticity measurements is attributed to the locally probed spin accumulation. The latter is in excellent agreement with simulations based on spin diffusion, valid for the 80 nm Cu spacer layer, together with spin sources given by the time derivative of the magnetization dynamics of Fig. 29.

The spin accumulation originating from the reference layer has a characteristic bipolar shape originating from the demagnetization and subsequent remagnetization of that same layer (Fig. 29 (c) and (e)). This temporal inversion of the spin accumulation provides a natural explanation for the bipolar switching of the free layer (Fig. 29 (b) and (d)). Starting from the antiparallel (AP) configuration, the first contribution arises from the demagnetization-driven spin current, which injects angular momentum favoring a temporary AP→P switching before the second contribution due to remagnetization of the reference layer finally favors P→AP switching. At higher fluence, the remagnetization of the reference layer is suppressed, allowing a permanent AP→P switching. When starting from the parallel (P) configuration, the spin accumulation generated during the remagnetization of the reference layer dominates and drives P→AP switching of the free layer. Importantly, the study shows that the experimentally measured ultrafast Kerr signal contains both magnetization and spin-accumulation contributions, and that separating these components is essential to correctly identify the origin, polarity, and timing of the spin currents involved. Together, these results establish that bipolar single-pulse all-optical switching in spin valves is governed by the time-resolved balance between demagnetization- and remagnetization-induced spin accumulation, rather than by a single, static spin-transfer process. But note that the mechanism of P to AP switching in ferromagnetic spin valves is still debated. While in spin valves with thick spacer layers, the reversal of the free layer comes from the remagnetization of the reference layer[137], the magnetization reversal for thin spacer layers may involve a "spin rotation" mechanism[132] or equivalently a reflected spin current dominated by minority spins [131]. Several theoretical frameworks have been adopted so far[132,312–315], but none of them is so far able to provide a complete and accurate picture capable of describing the very complex and non-equilibrium spin transport at play[316]. A similar but slower magnetization reversal in spinvalves is also possible not via spin currents but via RKKY coupling[138,139]. A detailed explanation of this phenomenon has been described in a recent review[140].

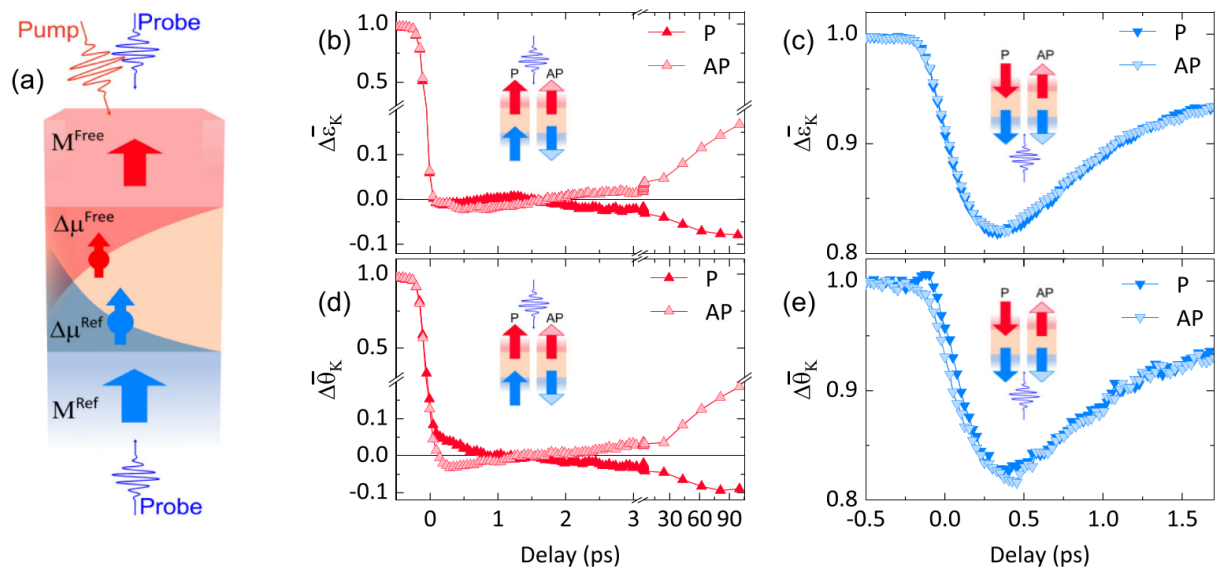

*Figure 29. Ultrafast spin accumulation and its role in bipolar all-optical switching of a ferromagnetic spin valve. Figure reproduced from Fig. 1 of Ref.[312]. (a) Schematic of the ferromagnetic spin-valve structure and pump-probe geometry, defining the free and reference layers separated by a Cu spacer. (b–e) Time-resolved magneto-optical Kerr ellipticity and rotation measured on the free and reference layers for parallel (P) and antiparallel (AP) magnetic configurations. Differences between Kerr rotation and ellipticity during the first picosecond reveal the presence of a transient spin accumulation in addition to the local magnetization dynamics.*

Finally, fully deterministic single-pulse switching in spin-valve structures was achieved by deliberately engineering the interplay between ultrafast laser-induced heating and non-local spin-current injection. In Ref.[317], Gd-free ferrimagnetic/ferromagnetic spin-valve heterostructures were designed such that a single femtosecond laser pulse can deterministically induce either antiparallel-to-parallel (AP→P) or parallel-to-antiparallel (P→AP) switching. As illustrated in Fig. 30, the final magnetic configuration is uniquely determined by a single laser parameter—either the pulse fluence or the pulse duration—and is completely independent of the initial magnetic state and the number of applied pulses. This behavior contrasts sharply with conventional toggle-type all-optical switching and establishes a true single-shot write-erase functionality.

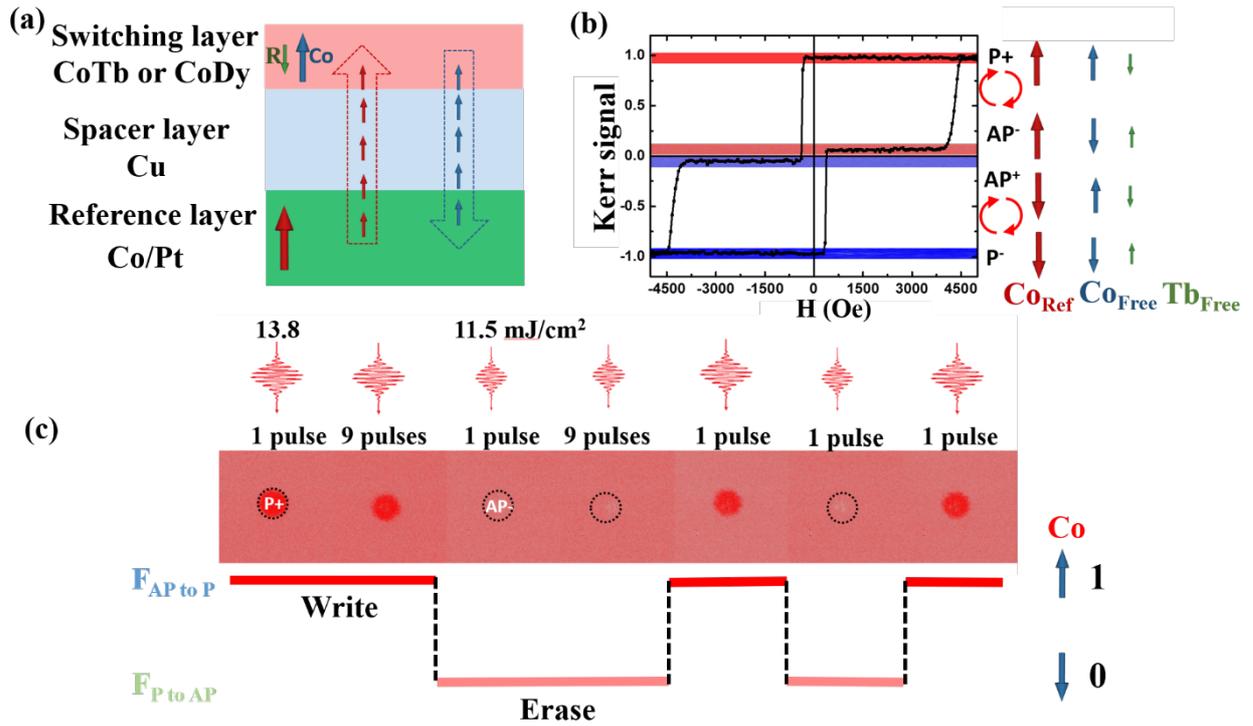

*Figure 30.* Deterministic single-pulse all-optical switching in a Gd-free ferrimagnetic spin-valve structure. Figure reproduced from Fig. 1 of Ref.[317]. *(a) Schematic representation of the ferrimagnetic/ferromagnetic spin-valve structure, consisting of a high-anisotropy ferromagnetic reference layer and a ferrimagnetic free layer separated by a Cu spacer, illustrating the generation of ultrafast spin currents following femtosecond laser excitation. (b) Polar magneto-optical Kerr hysteresis loop showing four stable magnetic configurations (P+, AP+, AP−, P−), evidencing magnetic decoupling between the layers. (c) Kerr microscopy images demonstrating deterministic magnetic writing and erasing using single femtosecond laser pulses. For a fixed pulse duration, a higher laser fluence deterministically sets the spin valve into the parallel state, while a lower fluence sets it into the antiparallel state, independently of the initial magnetic configuration and the number of applied pulses. This figure demonstrates true single-shot deterministic switching enabled by ultrafast spin-current-mediated angular-momentum transfer.*

The underlying physical mechanism relies on a controlled balance between ultrafast spin-current generation and transient thermal excitation of the ferrimagnetic free layer. As summarized by the state diagram in Fig. 31, P→AP and AP→P transitions occur in distinct regions of the fluence-pulse-duration parameter space, reflecting different regimes of demagnetization- and remagnetization-driven spin-current injection. In this architecture, the thermal stability of the stored information is governed by the perpendicular magnetic anisotropy of the free layer, while the energy required to write a bit is determined by the Curie temperature of the ferrimagnetic layer and the efficiency of ultrafast spin-current transfer. As a result, thermal stability and writing energy are no longer intrinsically coupled.

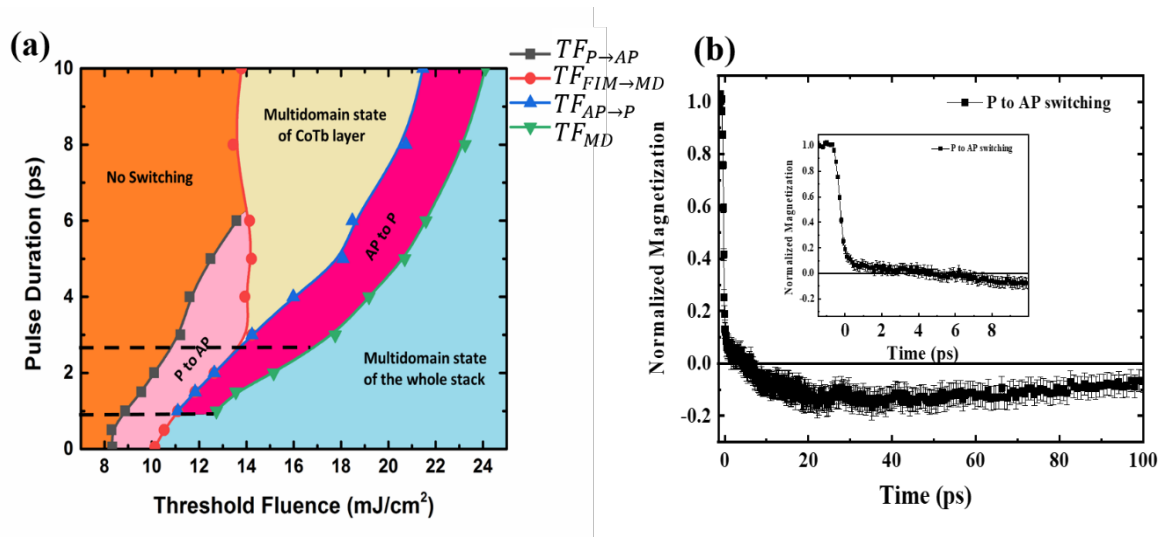

*Figure 31. State diagram and ultrafast switching regimes in a ferrimagnetic spin-valve system. Figure reproduced from Fig. 3 of Ref.[317]. (a) State diagram mapping the final magnetic configuration of the spin valve after a single femtosecond laser pulse as a function of pulse duration and laser fluence. Distinct regions corresponding to no switching, P→AP switching, AP→P switching, and multidomain formation are identified. The diagram highlights that deterministic switching between parallel and antiparallel states can be achieved either by tuning the laser fluence at fixed pulse duration or by adjusting the pulse duration at fixed fluence. (b) Time-resolved magneto-optical Kerr measurements of the ferrimagnetic free layer showing ultrafast demagnetization followed by magnetization reversal on a picosecond timescale. Together, these results demonstrate that the thermal stability of the stored information, set by the magnetic anisotropy of the free layer, is decoupled from the optical writing energy, which is governed by the Curie temperature and spin-current efficiency. This separation of thermal stability and write energy establishes a favorable framework for ultrafast, energy-efficient opto-spintronic memory devices*

This decoupling represents a fundamental advantage over field-driven or current-driven switching schemes, such as spin-transfer torque and spin-orbit torque, in which increasing thermal stability inevitably requires higher switching currents or fields, leading to increased energy dissipation and reduced scalability. In contrast, the spin-valve-based all-optical approach demonstrated by Zhang *et al.*[317] enables the use of high-anisotropy, thermally stable materials while maintaining low optical writing energies. This constitutes the first magnetic writing mechanism in which thermal stability and write energy are not competing requirements, establishing ultrafast spin-current-mediated all-optical switching as a highly promising route for future opto-spintronic memory devices.

The switching energy, typically a few femtojoules for a 20 × 20 nm² bit, is nearly an order of magnitude below that of STT-MRAM, illustrating the efficiency of non-thermal spin transport[131,135]. Both optical and electronic excitation follow a common sequence involving ultrafast electron heating, unequal sublattice demagnetization, rapid angular-momentum exchange, and deterministic reversal[130].

Recent experiments by Ishibashi et al.[136] further extended the concept of ultrafast spin-current-mediated switching in ferromagnetic spin valves by demonstrating that hot-electron pulses alone are sufficient to induce deterministic magnetization reversal, without requiring direct optical excitation of the magnetic layers. The studied system is a ferromagnetic spin valve of the same class as those discussed in Figs. 28 and 29, consisting of two perpendicularly magnetized Co/Pt multilayers acting as reference and free layers, separated by a Cu spacer. As in previous optical switching studies, the reference layer has a higher Curie temperature than the free layer, ensuring selective demagnetization of the free layer under ultrafast excitation.

Single-shot magneto-optical Kerr microscopy measurements reveal that a single ultrashort heat pulse can fully demagnetize the free layer and trigger deterministic parallel-to-antiparallel (P→AP) switching, even in the absence of direct optical absorption in the magnetic layers. The persistence of switching in this regime, highlighted in Fig. 32, unambiguously demonstrates that nonlocal electronic heat transport and the associated ultrafast spin-current generation are the essential driving mechanisms. Control experiments using insulating barriers further confirm that blocking electronic transport suppresses switching, emphasizing the central role of hot-electron-mediated angular-momentum transfer.

Importantly, this establishes that femtosecond electronic switching can be implemented within standard metallic multilayer stacks, without optical access to the active magnetic layer. As a consequence, this approach provides a realistic pathway toward CMOS-compatible ultrafast spintronic devices, where femtosecond electrical or thermal pulses could replace laser excitation while preserving the speed and efficiency of spin-current-driven switching.

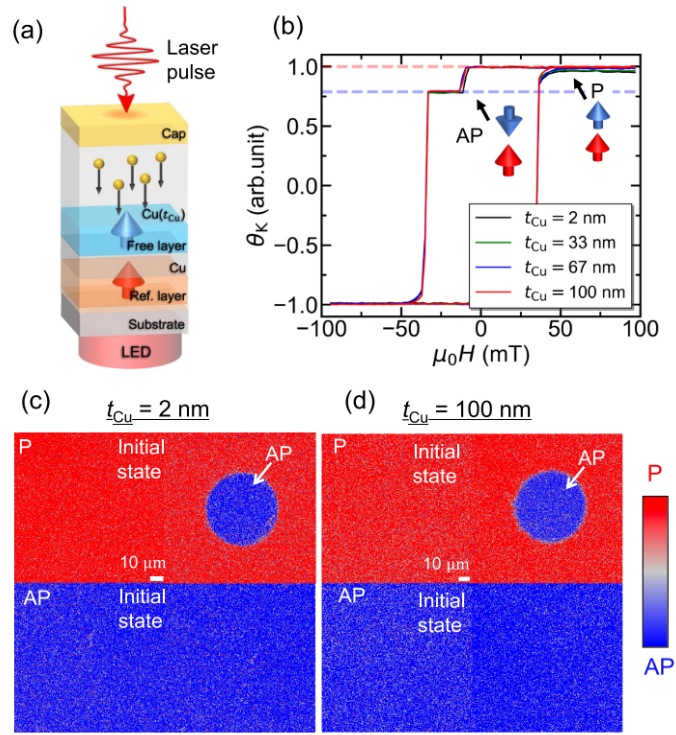

*Figure 32.* Single-shot magnetization reversal in a ferromagnetic spin valve driven by ultrafast heat transport. (a) Schematic of the Co/Pt-Cu-Co/Pt spin-valve stack capped with a wedged Cu layer of thickness $t_{Cu}$ and of the experimental geometry. The Cu overlayer controls the energy-deposition profile, from direct optical excitation of the magnetic layers to purely indirect excitation via hot-electron transport. (b) Polar MOKE hysteresis loops defining the parallel (P) and antiparallel (AP) magnetic states of the spin valve. (c,d) Single-shot MOKE images recorded before and after laser excitation for $t_{Cu} = 2$ nm and $t_{Cu} = 100$ nm, respectively. Deterministic P-to-AP switching of the free layer is observed even when the laser light is fully absorbed in the thick Cu and capping layers, demonstrating

*that ultrashort hot-electron pulses alone are sufficient to trigger magnetization reversal in ferromagnetic spin valves.*

**4.4 From spin valves to tunnel junctions**

For any realistic technological application, the ultrafast spin-transfer concepts demonstrated so far in metallic spin valves must ultimately be transposed to magnetic tunnel junctions (MTJs), which remain the cornerstone of MRAM technology owing to their large tunnel magnetoresistance (TMR) and scalable electrical readout. This transition is, however, non-trivial. In metallic spin valves, femtosecond demagnetization launches hot electrons that propagate ballistically or superdiffusively through metallic spacers, efficiently transferring angular momentum to the free layer and generating an ultrafast spin-transfer torque. In MTJs, by contrast, the presence of an MgO tunnel barrier strongly suppresses charge transport and filters the electronic population in both energy and symmetry, raising the fundamental question of whether femtosecond spin-transfer torque can survive tunneling on sub-picosecond timescales.

Recent experiments demonstrate that this is indeed the case. Spin-polarized hot electrons can tunnel through MgO barriers while retaining a remarkably high degree of spin polarization—exceeding 80%—thereby generating a sub-picosecond spin-transfer torque capable of reversing the magnetization of the free layer. Ultrafast magneto-transport and terahertz-emission measurements converge toward a unified picture in which demagnetization-driven hot electrons act as broadband spin injectors across metals, semiconductors, and insulating barriers alike[143]. Within this framework, the tunnel barrier no longer constitutes a fundamental limitation but rather an energy- and spin-selective filter whose properties can be engineered to optimize femtosecond spin injection.

This perspective is strongly reinforced by the work of Geiskopf *et al.*[318], who demonstrated single-shot, helicity-independent all-optical switching in in-plane-magnetized MTJs while preserving a TMR exceeding 100%. In these devices, a CoFeB/MgO/CoFeB junction is exchange-coupled to an optically switchable CoGd layer, allowing the ultrafast reversal of the ferromagnetic free layer to be monitored directly through the tunnel conductance. Despite the presence of an optically opaque top electrode that partially screens the excitation, deterministic toggle switching driven by a single femtosecond pulse is observed, highlighting the robustness of ultrafast angular-momentum transfer in technologically realistic MTJ stacks. Importantly, the in-plane geometry relaxes constraints associated with perpendicular anisotropy while remaining fully compatible with high-temperature annealing and standard MTJ processing.

Complementary experiments by Igarashi *et al.*[133] further extend this picture to rare-earth-free CoFeB/MgO-based MTJs, in which structural and thermal asymmetry is deliberately engineered via the capping layer. Figure 33 demonstrates that by controlling the laser-energy absorption profile, deterministic parallel-to-antiparallel switching can be achieved after a single femtosecond pulse. The reversal is directly visualized by Kerr microscopy and independently confirmed through tunnel magnetoresistance measurements in micro-fabricated devices. Optical simulations reveal that thicker Ru or Pt capping layers selectively reduce absorption in the reference layer while maintaining sufficient energy density in the free layer to induce complete demagnetization and the associated ultrafast spin imbalance across the MgO barrier. These results establish that femtosecond spin-transfer torque can be preserved across tunnel barriers, provided that interface transparency, barrier thickness, and optical absorption profiles are carefully optimized.

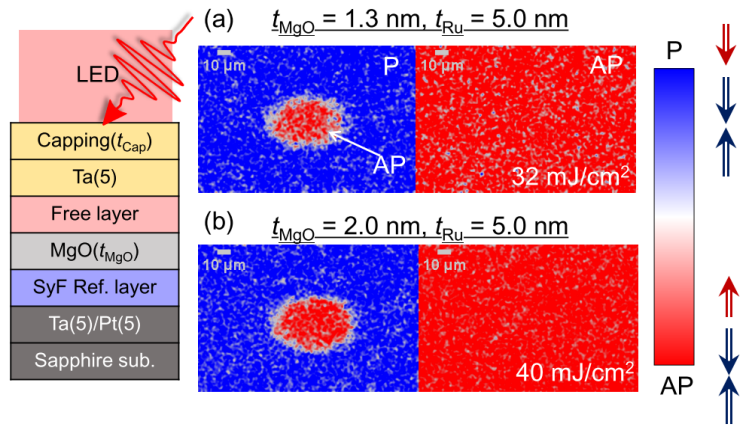

*Figure 33.* Single-shot all-optical switching in CoFeB/MgO magnetic tunnel junctions enabled by engineered optical absorption (adapted from Fig. 2 of Ref.[133]). (a) Kerr microscopy images showing deterministic parallel-to-antiparallel (P→AP) switching of the CoFeB free layer after a single femtosecond laser pulse, starting from the P (blue) and AP (red) magnetic configurations, for MgO-based MTJs with optimized Ru or Pt capping layers. (b) Corresponding switching behavior observed for different MgO thicknesses, demonstrating robustness against barrier thickness. (c) Calculated laser energy absorption in the capping, free, and reference layers as a function of capping-layer thickness. Increasing the Ru or Pt cap selectively suppresses absorption in the reference layer while maintaining sufficient energy density in the free layer to induce complete demagnetization and the associated ultrafast spin imbalance across the MgO barrier. (d) Electrical detection of the laser-induced reversal in micro-fabricated MTJ devices via tunnel magnetoresistance (TMR), confirming that the optically driven switching directly translates into a change of the tunnel conductance. Together, these results establish that deterministic femtosecond spin-transfer torque and ultrafast magnetization switching can be preserved across an insulating tunnel barrier, provided that the optical absorption profile and interface asymmetry are carefully engineered.

Looking forward, progress will rely on engineering MTJs in which MgO simultaneously ensures high-fidelity electrical readout and efficient femtosecond spin injection. Achieving this balance will open a realistic pathway toward CMOS-compatible ultrafast memory and logic devices operating at fundamentally new speed limits, where information is written optically, transported electronically, and processed magnetically on sub-picosecond timescales.

## 4.5 Outlook

The convergence of femtosecond optical excitation, hot-electron transport, and spin-current engineering now outlines a clear pathway toward hybrid photonic-spintronic devices operating at the ultimate temporal limits of magnetic order. Throughout this review, we have seen how ultrafast demagnetization, initially studied as a fundamental non-equilibrium phenomenon, naturally leads to the generation of intense femtosecond spin currents and to deterministic magnetization switching in a variety of magnetic systems. Translating these discoveries into functional technologies will require several key advances, including reproducible and deterministic switching at the nanoscale, seamless integration with silicon photonics and CMOS-compatible electronics, optimized interface transparency for efficient angular-momentum transfer, and robust thermal management under high-repetition-rate operation.

If deterministic magnetic switching can ultimately be achieved at sub-femtojoule energy scales, the long-standing separation between magnetic memory and photonic logic may begin to disappear. In such architectures, information could be written optically, transported electronically through spin-polarized carriers, and processed magnetically with unprecedented speed and energy efficiency. In this sense, the original vision of femtosecond magnetic logic first anticipated by Beaurepaire and collaborators[20] is gradually evolving from a conceptual idea into a realistic technological objective, as ultrafast magnetism increasingly develops into a field of engineering as well as fundamental physics.

Recent observations of electron-magnon scattering occurring on sub-10 fs timescales[63] further suggest that the intrinsic speed limits of spintronic devices may extend well beyond what was previously assumed. Although these measurements were obtained in the low-fluence regime and primarily associated with ultrafast demagnetization, there are compelling indications that such strong electron-magnon coupling may persist in the highly non-linear regime relevant for magnetization reversal[33,61,62,68]. Since electron-magnon scattering, often described within the framework of s-d exchange coupling, constitutes one of the fundamental interactions governing ultrafast angular-momentum transfer[130], it is conceivable that magnetization switching could ultimately approach the few-femtosecond timescale, provided that spin currents with comparable temporal bandwidth can be generated. In this perspective, intense mid-infrared excitation[319] appears as a particularly promising route for accessing this extreme non-equilibrium regime and pushing ultrafast spintronics toward its ultimate physical limits.

# 5. Conclusion

Over the past three decades, the discovery of femtosecond laser-induced ultrafast demagnetization has profoundly reshaped our understanding of angular-momentum transfer in magnetic materials. What initially emerged as a fundamental question in femtomagnetism has progressively evolved into the broader framework of ultrafast spintronics, in which optical excitation not only quenches magnetic order but also generates intense ultrashort spin currents capable of driving deterministic magnetization dynamics. A key insight emerging from this field is that the rapid collapse of magnetization acts as a powerful source of femtosecond spin currents. These currents can propagate across complex heterostructures and exert ultrafast spin-transfer torques on neighboring magnetic layers. The identification of microscopic channels ranging from spin-orbit-mediated scattering and electron-magnon interactions to superdiffusive transport and interfacial spin injection has revealed that angular momentum can be redistributed across magnetic multilayers on sub-picosecond timescales.

One of the central consequences of this mechanism is the realization that demagnetization-driven spin currents represent an ultrafast analogue of conventional spin-transfer torque, operating several orders of magnitude faster than current-driven switching in conventional spintronic devices and at extremely low energy scales. This principle has enabled deterministic single-pulse switching not only in rare-earth-transition-metal ferrimagnets but also in fully ferromagnetic spin valves and magnetic tunnel junctions. These demonstrations remove the need for specialized ferrimagnetic materials and open new pathways toward technologically relevant device architectures based on ultrafast spin transport.

More broadly, the convergence of all-optical switching, ultrafast spin-transfer torque, and tunnel-junction readout establishes a unified physical picture in which optical, electronic, and spin degrees of freedom become tightly intertwined in a strongly non-equilibrium regime. In this emerging framework, femtosecond optical excitation provides the energy input, hot-electron transport redistributes angular momentum across interfaces, and magnetic heterostructures convert these spin currents into deterministic switching events. Beyond individual device demonstrations, recent advances in racetrack memories, integrated photonics, and hot-electron-driven switching suggest that ultrafast spintronics can be integrated into scalable and CMOS-compatible platforms. These developments point toward hybrid photonic-spintronic technologies in which information is written optically, transported magnetically, and read electrically, combining the speed of photonics with the non-volatility of magnetic storage. At the same time, these advances raise fundamental questions concerning the limits of angular-momentum transfer, the role of correlations and coherence in non-equilibrium spin systems, and the ultimate speed-energy trade-offs achievable in magnetic devices. Addressing these questions will require continued progress in both experimental techniques and theoretical descriptions of strongly non-equilibrium magnetism.

Ultimately, the developments reviewed here reveal a deeper paradigm shift: magnetization is no longer viewed as a slow thermodynamic variable but as a dynamic quantity that can be manipulated on femtosecond timescales through controlled flows of angular momentum. By bridging femtomagnetism and spintronics, ultrafast spintronics is transforming our ability to control magnetic order and opening the door to a new generation of information technologies operating at the ultimate temporal and energetic limits of magnetic matter.

Acknowledgments

This work was supported by the Agence Nationale de la Recherche (ANR) through the SLAM project (ANR-23-CE30-0047), and by the France 2030 government grants PEPR Electronics EMCOM (ANR-22-PEEL-0009), PEPR SPIN Chirex (ANR-22-EXSP-0002), SPINMAT (ANR-22-EXSP-0007), and OptoSpinCom (ANR-24-EXSP-0010), as well as by the MAT-PULSE Lorraine Université d'Excellence project (ANR-15-IDEX-04-LUE). This article is also based upon work carried out within COST Actions CA17123 MAGNETOFON and CA23136 CHIROMAG, supported by COST (European Cooperation in Science and Technology). Further support from the European Union's Horizon 2020 Research and Innovation Programme under the Marie Skłodowska-Curie Grant Agreement No. 861300 (COMRAD) is gratefully acknowledged. The authors also acknowledge support from the Institut Carnot ICEEL, the Région Grand Est, and the Métropole du Grand Nancy through the Fastness, OPTIMAG, and Opti-Memory projects, as well as from a Royal Society Research Fellowship and a Leverhulme Trust Research Project Grant (RPG-2023-271). Experiments at Institut Jean Lamour were performed using equipment from CC DAUM, CC Minalor, CC MagCryo

The authors warmly thank all the colleagues with whom they have had the pleasure to collaborate and to share many fruitful and stimulating discussions. In particular, they acknowledge Michel Hehn, Grégory Malinowski, Julius Hohlfeld, Jon Gorchon, Thomas Hauet, Daniel Lacour, Sebastien Petit-Watelot, Juan Carlos Rojas-Sanchez, Yuan Lu, Francois Montaigne, Charles-Henri Lambert, Matthias Gottwald, Mohamed El Hadri, Philippe Scheid, Marwan Deb, Georgy Kichin, Junta Igarashi, Jun-Xiao Lin, Zongxia Guo, Kaushalya Jhuria, Danny Petty Gweha Nyoma, Jean Loïs Bello, Maxime Vergès, Alberto Anadón Barcelona, Harjinder Singh, Yann Le Guen, Boonthum Kunyangyuen, Aleena Joseph, Matthias Riepp, Corentin Aulagnet, Jude Compton Stewart, Pierre Chailloleau, Victor Huet at Université de Lorraine; Tobias Kampfrath, Reza Rouzegar, Oliver Gückstock and Alexander Chekhov, Junwei Tong, Martin Weinelt, Piet Brouwer, Oliver Franke and Gal Lemut at Freie Universität